\documentclass[conference]{IEEEtran}
\IEEEoverridecommandlockouts
\usepackage{cite}

\usepackage{amsmath,amssymb,amsfonts}
\usepackage{amsthm}
\usepackage{algorithmic}
\usepackage{graphicx}
\usepackage{textcomp}
\usepackage{xcolor}
\usepackage[font=small,skip=2pt]{caption}
\usepackage{soul}
\usepackage{verbatim}
\usepackage{enumitem}
\usepackage[normalem]{ulem}
\usepackage[hidelinks]{hyperref}
\AtBeginDocument{\def\sectionautorefname{Section}}%

\usepackage{dblfloatfix}    

\usepackage{tikz}
\newcommand*\circled[1]{\tikz[baseline=(char.base)]{
            \node[shape=circle,draw,inner sep=1pt,font=\sffamily\footnotesize] (char) {\textbf{#1}};}}
\usepackage{subcaption}

\theoremstyle{definition}

\newcommand{\lmttfont}{\fontfamily{lmtt}\selectfont}

\newcommand{\toolname}[1]{\emph{IRIS}}
\newcommand{\fuzzername}[1]{\emph{NOME\_FUZZER}}

\usepackage{pdfcomment}
\usepackage{todonotes}
\newcommand{\revision}[1]{{\textcolor{black} {#1}}}

\def\BibTeX{{\rm B\kern-.05em{\sc i\kern-.025em b}\kern-.08em
    T\kern-.1667em\lower.7ex\hbox{E}\kern-.125emX}}

\begin{document}


\title{IRIS: a Record and Replay Framework to Enable Hardware-assisted Virtualization Fuzzing}


\author{
\IEEEauthorblockN{Carmine Cesarano, Marcello Cinque, Domenico Cotroneo, Luigi De Simone, Giorgio Farina}
\IEEEauthorblockA{\textit{Universit\`a degli Studi di Napoli Federico II, Italy} \\  
\{carmine.cesarano, macinque, cotroneo, luigi.desimone, giorgio.farina\}@unina.it}
}


\maketitle

\begin{abstract}

Nowadays, industries are looking into virtualization as an effective means to build safe applications, thanks to the isolation it can provide among virtual machines (VMs) running on the same hardware. In this context, a fundamental issue is understanding to what extent the isolation is guaranteed, despite possible (or induced) problems in the virtualization mechanisms.
Uncovering such isolation issues is still an open challenge, especially for \revision{hardware-assisted virtualization, since the search space should include all the possible VM states (and the linked hypervisor state)}, which is prohibitive. In this paper, we propose \textit{IRIS}, a framework \revision{to record (learn)  sequences of inputs  (i.e., VM seeds) from the real guest execution (e.g., OS boot), replay them as-is to reach valid and complex VM states, and finally use them as valid seed to be mutated for enabling fuzzing solutions for hardware-assisted hypervisors.} We demonstrate the accuracy and efficiency of \textit{IRIS} in automatically reproducing valid VM behaviors, with no need to execute guest workloads. We also provide a proof-of-concept fuzzer, based on the proposed architecture, showing its potential on the Xen hypervisor.

\end{abstract}


\section{Introduction}



The role of virtualization technology is widening today, from cloud computing systems to critical industrial systems in several domains (e.g., railways, avionic, automotive) \cite{cinque2021virtualizing}, due to its ability to reduce \textit{SWaP-C} factors (size, weight, power, and cost) by consolidating multiple software stacks on the same system-on-a-chip (SoC)\cite{cinque2021virtualizing, blackberry}.

\revision{The \textit{isolation} properties of virtualization \cite{Popek1973FormalRF} are a key factor for its use as an enabling technology for \textit{Industry 4.0} \cite{cinque2021virtualizing, cilardo2022virtualization}. Isolation properties play also a key role in functional safety standards (e.g., DO178C for avionic \cite{do178C}, ISO 26262 for automotive \cite{iso26262}, etc.), which recommend providing evidence on temporal, memory, and fault isolation among applications, sharing the same computing infrastructure. The violation of such properties can lead to catastrophic consequences \cite{lezzi2018cybersecurity}. }

\revision{Virtualization allows dividing the hardware resources of a computer into multiple \textit{virtual} computers, called Virtual Machines (VMs) or \textit{guests}. It is implemented} by a software layer, the \textit{hypervisor}, which abstracts physical CPU, memory, and I/O, to run different and isolated application environments, including the operating system (OS). 
In particular, \textit{hardware-assisted virtualization} takes advantage of CPU virtualization technologies (e.g., Intel VT-x \cite{neiger2006vtx}, AMD-V \cite{amd_svm}, ARM VHE \cite{dall2016arm}) to implement  \textit{full virtualization}, i.e., emulating a complete machine to run unmodified guests.
These hardware extensions \revision{introduce a novel (and higher) level of privilege for the hypervisor. This way, developers can implement} \textit{virtual CPU (vCPU)} abstractions that can run a guest OS at a lower level of privilege compared to the hypervisor. 
\revision{If the guest OS needs to execute a \textit{sensitive} instruction (e.g., page table update, interrupt handling, etc.), the execution traps, and the control is passed to the hypervisor. The switch from the guest OS to the hypervisor is called \textit{VM exit}. A VM exit also involves a change in the privilege level, from the VM to the hypervisor's most privileged mode. This passage, however, }
can put isolation at risk \cite{CVE-2020-2732, CVE-2011-1936, CVE-2010-0435}, \revision{possibly} leading to hypervisor crashes (together with all running VMs) or denial of service issues (hangs, low responsiveness, etc.). 


Uncovering such \textit{isolation issues} is a major concern and a compelling open challenge to foster the adoption of hypervisors in critical industrial domains. This is especially true for hardware-assisted virtualization, as the search \revision{for isolation issues} should include all the possible VM states reached \revision{by an extremely high number of combinations of CPU instructions}. 

In the literature, the problem is being tackled by \textit{fuzz testing}, which has proven to be very effective for security isolation assessment since it can reveal new vulnerabilities and bugs in complex software systems, including hypervisors \cite{zhu2022fuzzing}. The common fuzzing loop for hypervisors is implemented by
\revision{i) submitting a sequence of VM operations (or reverting a VM snapshot), to bring the VM into a given state (and its corresponding hypervisor state) and then ii) submitting fuzzing input (the corrupted \textit{seed}) to the hypervisor from the reached VM state.}


\revision{Existing hypervisor fuzzing solutions fall short when targeting hardware-assisted virtualization, for the following reasons.} 
First, state-of-the-art fuzzers mostly target I/O virtualization \cite{Schumilo2020HYPERCUBEHH, Myung2022MundoFuzzHF, Henderson2017VDFTE, Pan2021VShuttleSA, Bulekov2022MorphuzzB}, \revision{starting from the same VM state, thus } leaving behind the vCPU abstraction, which is the core of hardware-assisted hypervisors. 
Second, current studies are facing serious issues related to seed generation, \revision{which range from the building of a valid VM state to start the fuzzing \cite{Fonseca2018MultiNyxAM, Yan2018FastPS} to the generation of valid seeds to be corrupted to accelerate the fuzzing, which may require}
the use of an ad-hoc OS within VM \cite{Schumilo2020HYPERCUBEHH, Schumilo2021NyxGH}, up to manually constructed seeds \cite{Amit2015VirtualCV, Yan2018FastPS, Fonseca2018MultiNyxAM, Ge2021HyperFuzzerAE}. 
Each of these approaches includes a non-negligible effort, which heavily impacts fuzzing effectiveness, \revision{in terms of developers' and testers' work overhead}. 
\revision{Last, reaching a new VM state (e.g., setting a new CPU physical state) requires a deep knowledge of the underlying hardware (i.e., the knowledge of an operating system). Trying to reach valid VM states from a dumb sequence of inputs risks incur in several crashes of the test VM \revision{\cite{zhu2022fuzzing}}, which then requires resetting the test, with a non-negligible impact on the testing time efficiency}.

To overcome the above limitations, we propose \toolname{}\footnote{From Greek mythology, a messenger of the gods and the personification of the rainbow, which connects the world of the gods with humanity}, a framework \revision{to \textit{record} (learn)  sequences of inputs  (i.e., VM seeds) from the real guest execution (e.g., OS boot),  \textit{replay} them as-is to reach valid and complex VM states and finally use them as valid seed for a fuzzing.}

The current implementation of \toolname{} targets Xen (hardware-assisted mode) as a hypervisor, which is one of the most complex and popular virtualization solutions used today, also for industrial scenarios \cite{Schulz2022Evaluation, alonso2020analysing, sabogal2018towards}. Moreover, we target \textit{Intel VT-x} extensions due to the great support included for the majority of existing hardware-assisted hypervisors.

\revision{Key findings on a thorough experimental analysis on \toolname{} reveal:}

\begin{itemize}
    \item \revision{the \textit{accuracy} of \textit{IRIS} to automatically generate (record) seeds to reproduce real VM behaviors. The results show a fitting of code coverage ranging between $92.1\%$ and $100\%$ in our settings, compared to real guest execution;}
    
    \item \revision{the \textit{efficiency} of \textit{IRIS} to replay recorded seeds to reach a valid VM state, with a time improvement from $42.5\%$ to $99.6\%$ compared to the real guest execution;}
    
    
   

\end{itemize}

\revision{A preliminary implementation of a proof-of-concept fuzzer, built upon the findings above, demonstrates the feasibility of using the proposed approach as a first step for assessing hardware-assisted virtualization}.


We publicly released \toolname{}, raw data about our experiments and scripts to reproduce obtained results, and the proof-of-concept fuzzer\footnote{\url{https://github.com/dessertlab/iris}}. 

The rest of the paper is structured as follows. \sectionautorefname{}~\ref{sec:background} provides background on \revision{virtualization and the hardware support by Intel VT-x}. \sectionautorefname{}~\ref{sec:probl_statement} highlights the need for a record and replay framework to train semantically complex guest VM behaviors for fuzzing purposes. \sectionautorefname{}~\ref{sec:architecture} and \sectionautorefname{}~\ref{sec:xen_implementation} provides \toolname{} framework design and Xen implementation details, respectively. \sectionautorefname{}~\ref{sec:evaluation} provides a thorough experimental evaluation of effectiveness and efficiency of \toolname{}, as well as, the description of proof-of-concept fuzzer. \sectionautorefname{}~\ref{sec:discussion} discusses limitations and future works. \sectionautorefname{}~\ref{sec:related} provides the related work, and \sectionautorefname{}~\ref{sec:conclusion} concludes the paper.


\section{Background}
\label{sec:background}

\begin{figure*}[h]
    \centerline{\includegraphics[width=2\columnwidth]{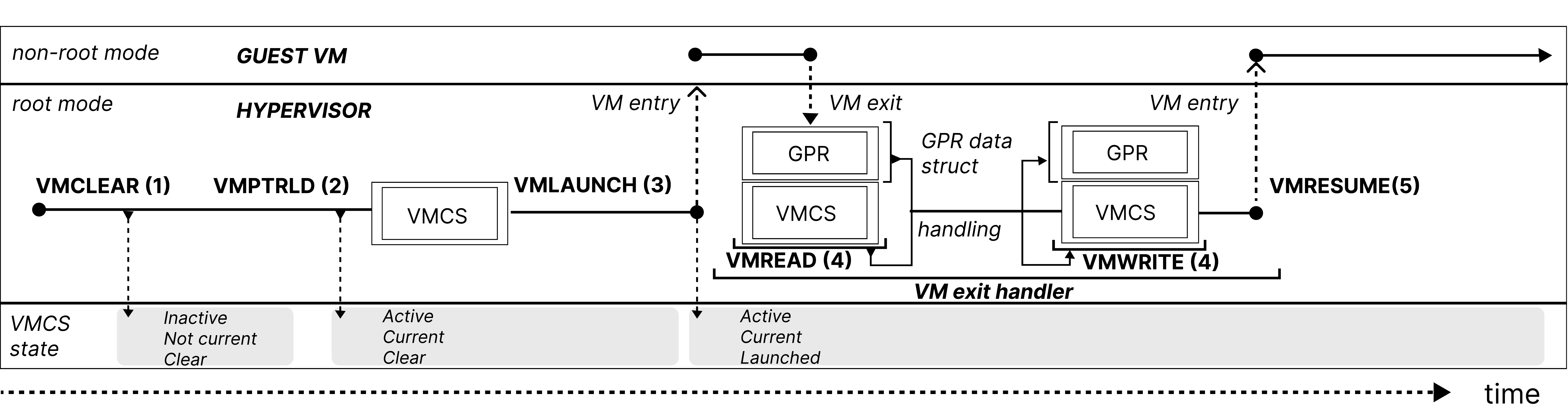}}
    \caption{Workflow of a virtual machine in VTX \label{fig:workflow_vmcs}}
    \label{fig}
\end{figure*}

\textbf{Virtualization}. Virtualization techniques allow running multiple Operating Systems (OSes) on the same hardware. The main software component that enables virtualization is the \textit{hypervisor} or \textit{Virtual Machine Monitor (VMM)}. It is responsible for creating, managing, and scheduling Virtual Machines (VMs), which represent an abstraction of CPUs, memory areas, and devices of a real machine. Early virtualization techniques were based on \emph{full virtualization}, which completely emulate privileged instructions and I/O operations. 
To achieve higher performance, \emph{paravirtualization} involves modifying guest OSes (i.e., OSes running as VMs) to directly use the services offered by the hypervisor through the so-called \emph{hypercalls}. In the last years, virtualization evolved to exploit hardware extensions in modern CPUs, to enable full virtualization while improving overall performance \revision{{\cite{neiger2006intel, adams2006comparison}}}. 

\revision{\textbf{Intel VT-x and VM Life-cycle.}
Intel VT-x} \cite{intel2009intel}, is the target technology we used to implement the record and replay framework. Once the virtualization is enabled ({\lmttfont VMXON} instruction), two operating modes are active. The hypervisor (\textit{VMM}) operates in \textit{root mode}, while the guest \textit{VM}s run in \textit{non-root mode}. The latter modes are orthogonal to traditional execution modes (long, protected, and real modes) and to privilege levels (i.e., rings). Running a new VM in \textit{non-root mode} requires allocating and initializing in memory a particular control structure, called \textit{Virtual Machine Control Structure (VMCS)}, linked to a specific vCPU. The VMCS, except for its first eight bytes, must be read and written by executing dedicated \textit{VMX instructions} called {\lmttfont VMREAD} and {\lmttfont VMWRITE}, otherwise unpredictable failure modes can occur (see Section 24.11.1 in \cite{intel2009intel}). The VMCS consists of the following areas: \textit{guest-state}, \textit{host-state}, \textit{control fields}, and \textit{VM exit information}. The first two are the most important in the context of our framework and include, respectively, the processor state when the VM is suspended and resumed. Specifically, they include special-purpose registers (e.g., control registers, instruction pointers, etc.). 

\figureautorefname{}~\ref{fig:workflow_vmcs} depicts the VM lifecycle. The VMCS is initialized ({\lmttfont VMCLEAR} instruction, step \circled{1} in \figureautorefname{}~\ref{fig:workflow_vmcs}) during the VM startup and subsequently loaded ({\lmttfont VMPTRLD} instruction, step \circled{2} in \figureautorefname{}~\ref{fig:workflow_vmcs}). When the VMCS is loaded, its internal hardware state becomes \textit{Active Current Clear}. In this state, the hypervisor can set up the VM, for example, by defining the events and instructions in \textit{non-root mode} that will cause a switch to the \textit{root mode} (i.e., a \textit{VM exit}). Once the setup is completed, the hypervisor can launch the VM ({\lmttfont VMLAUNCH} instruction, step \circled{3} in \figureautorefname{}~\ref{fig:workflow_vmcs}). \revision{Once this instruction is complete, the VMCS state becomes \textit{Active Current Launched} and the Guest VM can run, after switching to non-root mode (called \textit{VM entry}).}

\revision{During the execution of the VM, the control can pass to the hypervisor every time a \textit{VM exit} occurs, requiring a context switch from non-root to root mode. 
VM exits can occur for different reasons. }
Currently, Intel \textit{x86} architecture support 69 \textit{VM exit reasons} (Appendix C, table 1-c \cite{intel2009intel}). Most of them are due to the execution of sensitive instructions by the VM, such as {\lmttfont RDMSR}, {\lmttfont WRMSR}, or {\lmttfont CRx ACCESS}. Others include VM events or conditions to be handled by the hypervisor, such as triple fault, interrupts, and I/O port access. Finally, the hypervisor can decide to trap some VM conditions to follow the VM evolution (e.g., VM introspection \cite{Hebbal2015VirtualMI}) or to take scheduling and resource-sharing decisions (e.g., memory deduplication \cite{xenDeduplication}).

\revision{The VM exit is a key operation since it can be exploited to compromise the isolation properties of the hypervisor. Hence, we use it as the mean to submit a seed to the hypervisor and to test its operation. Let us analyze in detail the steps occurring from the VM exit up to the successive VM entry (VM resume), including the execution of the \textit{VM exit handler} in the hypervisor (steps \circled{4} and \circled{5} in \figureautorefname{}~\ref{fig:workflow_vmcs}).}
\revision{The VM exit requires a hardware context switch from non-root to root mode, that entails: (i) to save the physical processor state } in the guest-state area of the VMCS (except for general purpose registers (GPRs), saved in the hypervisor data structure), \revision{(ii) to load the new root mode processor state from the host-state area of the VMCS, including also the instruction pointer register (RIP), containing the start address of the VM exit handler.
After the context switch, the} VM exit handler identifies, from the VMCS, the cause of the exit and appropriately resolves it. 
\revision{More importantly, during the execution, the VM exit handler can access the entire VMCS ({\lmttfont VMREAD}, step \circled{4} in \figureautorefname{}~\ref{fig:workflow_vmcs}), hence its control flow depends} on VMCS fields.
Additionally, the VM exit handler can change the VM state in the VMCS (guest-state area) ({\lmttfont VMWRITE}, step \circled{4} in \figureautorefname{}~\ref{fig:workflow_vmcs}). Once the VM is resumed ({\lmttfont VMRESUME} instruction, step \circled{5} in \figureautorefname{}~\ref{fig:workflow_vmcs}), the new VM state becomes operational on the physical CPU. The VMRESUME performs a new \revision{(inverse)} hardware context switch, where the processor state is loaded from the guest-state area of the VMCS.

\section{Problem statement}
\label{sec:probl_statement}


\revision{In hardware-assisted virtualization architectures, the hypervisor code is mainly run when \textit{VM exits} occur, as described in the previous section. Thus, a fuzzer could \textit{simply} submit a sequence of VM exits to the hypervisor, to test its behavior. However, submitting even a single VM exit is not trivial, since the behavior of the VM exit handler depends on the contents of the VMCS and, thus, on the previous exits handled by the hypervisor.} 

\begin{figure}[!b]
    \centerline{\includegraphics[width=.75\columnwidth]{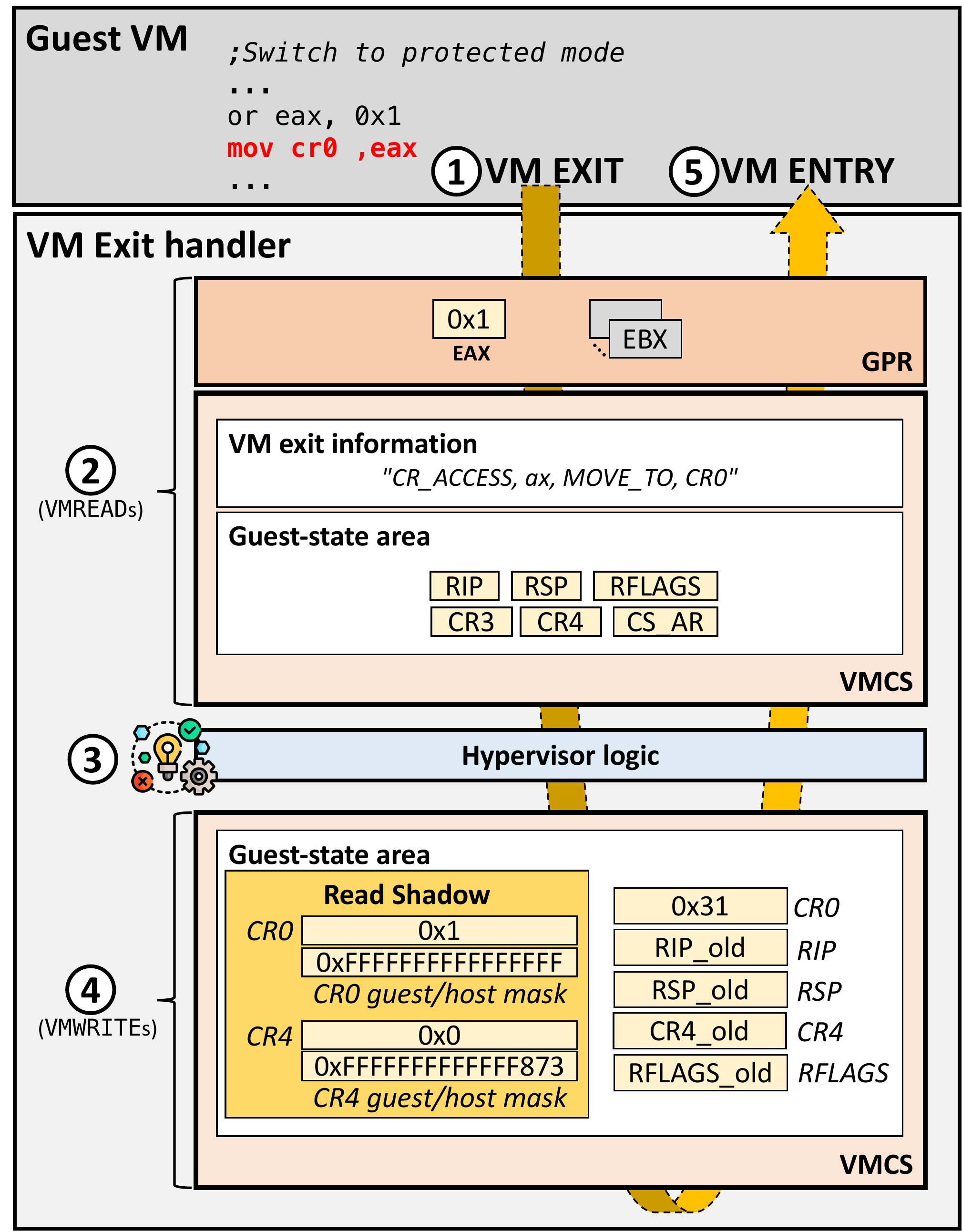}}
    \caption{Xen hypervisor control flow during switching into protected mode requested by a guest VM.}
    \label{fig:cr_example_hyp_actions}
\end{figure}

\revision{Let us substantiate this claim with an example.} Without loss of generality, we base \revision{the example} on the Xen hypervisor \revision{since it has a well-known implementation support of hardware-assisted Intel x86 virtualization \cite{barham2003xen}}. 
A frequent cause of exit is guest OS switch between different \revision{non-root} CPU modes, e.g., from \textit{real mode} to \textit{protected mode}. Switching to \textit{protected mode} (see Section 9.9.1 in \cite{intel2009intel} for details) requires numerous and complex preliminary operations, such as clearing interrupts, allocating the \textit{Global Descriptor Table (GDT)}, etc. Such operations must be performed in a strict order, otherwise, hardware exceptions will be triggered. 
\revision{The hypervisor intervenes several times during these operations, as some of the instructions executed by the guest are sensitive.}
Eventually, the guest VM tries to set the bit $0$ of the control register {\lmttfont CR0} equal to $1$ to enable \textit{protected mode}. \revision{Also this operation is sensitive, and thus it causes another VM exit} to the hypervisor (\revision{the one depicted as} step \circled{1} in 
\figureautorefname{}~\ref{fig:cr_example_hyp_actions}), \revision{with a given reason} (i.e., \textit{Control-register accesses}). 

\revision{Both guest GPR and VMCS fields are then read or written, during the handling of the required {\lmttfont CR0} access. The whole process is depicted in  \figureautorefname{}~\ref{fig:cr_example_hyp_actions}}. 
In particular, the hypervisor reads the \textit{VM exit information area} within the VMCS, but also other critical information \revision{from} the \textit{guest-state area} of the VMCS (step \circled{2} in 
\figureautorefname{}~\ref{fig:cr_example_hyp_actions}). The values read and the hypervisor's internal variables affect the VM exit handling performed by the hypervisor. At the end of the VM exit handling, the hypervisor updates its internal variables (e.g., including the abstraction of the current guest CPU operating mode) and writes the VMCS to update the state of the guest \revision{(steps \circled{3} and \circled{4} in 
\figureautorefname{}~\ref{fig:cr_example_hyp_actions})}.  
To make the picture even more complex, the access to control registers as {\lmttfont CR0} and {\lmttfont CR4} is regulated by the \textit{guest/host masks} and \textit{read shadows} during VM exit handling triggered by the \textit{MOVE to/from CR} operations. Finally, the control returns to the guest VM (step \circled{5} in 
\figureautorefname{}~\ref{fig:cr_example_hyp_actions}). 
{\revision{After the {\lmttfont CR0}} bit $0$ is finally set, if} the subsequent VM exits are successfully handled, the hypervisor updates the real value of {\lmttfont CR0}. 

\revision{Replicating} such a scenario from a \revision{"test" guest VM is prohibitive since it would mean writing the same logic, in terms of low-level instructions, like an operating system \cite{Schumilo2020HYPERCUBEHH, wei2019hyperbench}}. \revision{Even reproducing only the VM exit depicted in \figureautorefname{}~\ref{fig:cr_example_hyp_actions} would require to do all the preliminary exits to bring the hypervisor to a meaningful state for the target exit. } Instead, we propose a record and replay framework that allows to easily \textit{record} (learn) such a scenario, by executing a guest; further, the \textit{replay} mechanism allows submitting recorded VM behaviors as a train of recorded VM exits, with no need to execute guest workloads. 

In the following, \revision{we briefly highlight the main issues we have addressed by designing the} \toolname{} framework.

\begin{enumerate}[]
    
    \item \textbf{Automatic VM seed generation}: reduce the manual effort in VM seed generation, to minimize the required knowledge to develop a fuzzer for hardware-assisted hypervisors;
    
    \item \textbf{\revision{Accurate} VM seeds generation}: generate VM seeds that are \revision{accurate in reproducing the} real VM behaviors;
    
    \item \textbf{Efficient submission of VM seeds}: submit VM seeds in a lightweight fashion, without requiring specific VM guest workload to be executed.
    
\end{enumerate}

\section{IRIS Framework}
\label{sec:architecture}

\begin{figure}[h]
    \centerline{\includegraphics[width=.85\columnwidth]{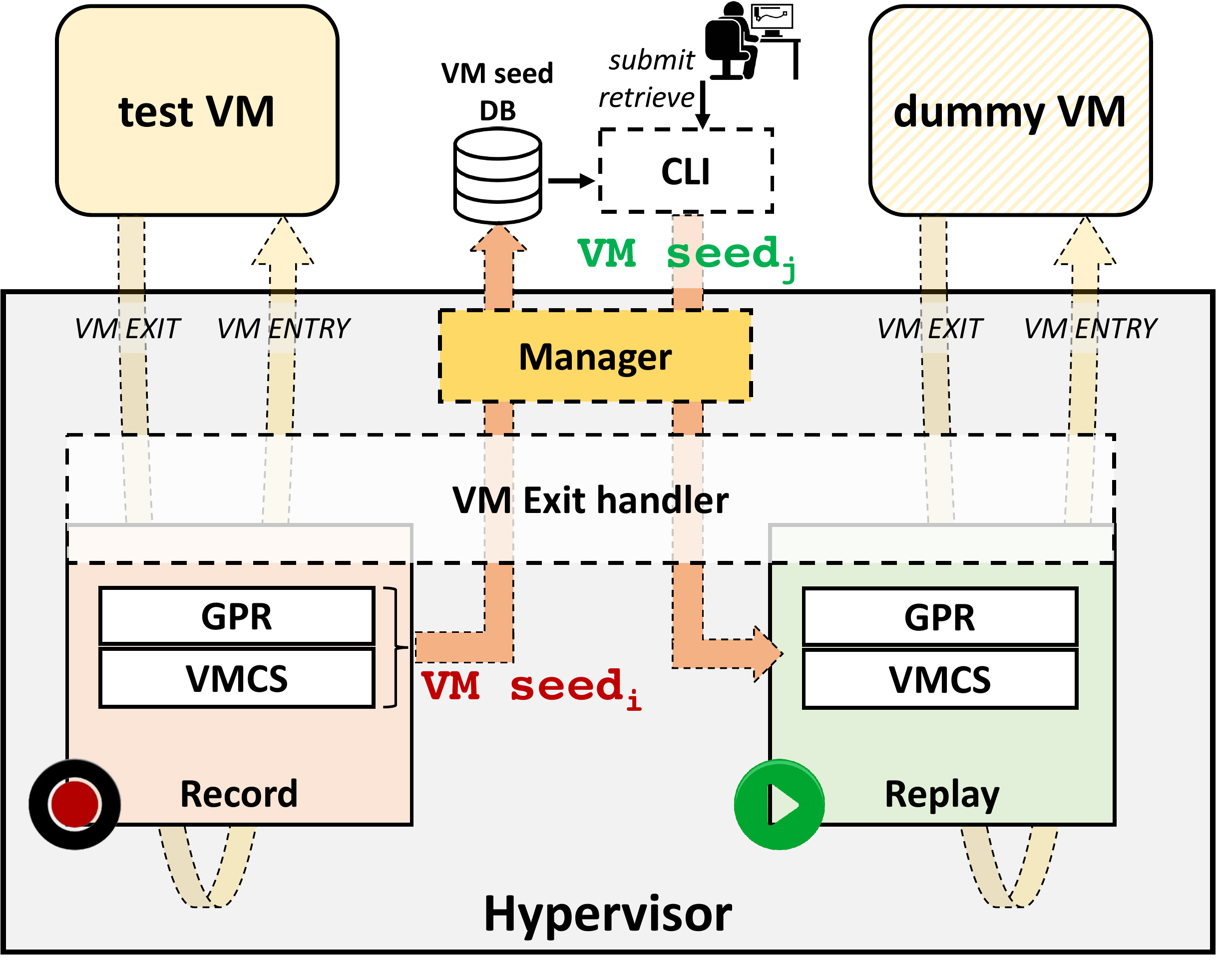}}
    \caption{Overview of \toolname{} design \label{fig:arch_abstract}}
    \label{fig}
\end{figure}

\revision{\figureautorefname{}~\ref{fig:arch_abstract} depicts the high level architecture of \toolname{}. 
The framework has been designed regarding Intel VT-x hardware-assisted virtualization technology and it does not strictly depend on the hypervisor under test.}
{\revision{In the description of \toolname{}, we use the following terminology:
\begin{itemize}    
    \item The \textit{\textbf{VM behavior}} is a sequence $VM\_exit\_trace =\{VM exit_{1}, ..., VM exit_{N}\}$ that is the flow of \textit{VM exits} triggered by a workload to reach a valid VM state;
    \item The \textit{\textbf{VM seed}} includes the pairs of VMCS \{field, value\} read via {\lmttfont VMREAD} instructions, and the values of general-purpose registers (GPR), both obtained during the handling of a VM exit within the \textit{VM\_exit\_trace}.
\end{itemize}
}}

In the following, we describe the architectural components that address the issues introduced in \sectionautorefname{}~\ref{sec:probl_statement}.

\subsection{Record}
\label{subsec:arch_recording}


The recording component aims to collect a set of information observed \revision{while executing the VM}, i.e., the \textit{VM behavior}. 
For each VM exit in a \textit{VM\_exit\_trace}, which is qualified by an \textit{exit reason}, the recording architecture stores \textit{(i)} the \textit{VM seed} and \textit{(ii)} \textit{metrics}, \revision{at hypervisor level.}


The \textit{metrics} are necessary to assess the \revision{accuracy and efficiency} of our framework in replaying the recorded VM behavior. Specifically, the \toolname{} framework records the \textit{i)} code coverage at hypervisor level, \textit{ii)} the pairs of VMCS \{field, value\} written via {\lmttfont VMWRITE} instructions, and \textit{iii)} and time needed during a VM exit. Code coverage at the hypervisor level is the simplest metric to estimate how much replaying \textit{VM seeds} is \textit{accurate} \revision{(refer to \subsectionautorefname{}~\ref{subsec:arch_replayig}, in which we define what is an accurate replay)} compared to the recorded VM behavior. 
\revision{However, coverage-related metrics could not always guarantee to cover critical areas that describe the VM behavior during execution. To mitigate this point, we also included in the \textit{IRIS} framework the monitoring of VMCS {field, value} pairs accesses, which are peculiar to the hardware-assisted hypervisor solutions. Specifically,} we use {\lmttfont VMWRITE}s for a more fine-grained validation of actual VM state changes from the point of view of the VMCS and VMX operations. Finally, the recording component (as well as the replaying, see \subsectionautorefname{}~\ref{subsec:arch_replayig}) gathers time needed to VM exit handling (via \revision{\textit{CPU cycle counters}}) as an \revision{efficiency} indicator.

\revision{Note that the recording mechanism implemented in \textit{IRIS} allows us to monitor whatever VM exit and related VMCS data/control information (i.e., VMREAD and VMWRITE operations). We deliberately avoid recording the \textit{test VM} memory as the only part of the VM state that we do not monitor. This choice was made to reduce the complexity of VM seeds compared to snapshot-based approaches {\cite{Ge2021HyperFuzzerAE, Schumilo2020HYPERCUBEHH}}. In \sectionautorefname{}~\ref{sec:evaluation}, we demonstrate that this choice is a good compromise to accurately record VM behaviors, and we discuss the few cases in which IRIS can not obtain good accuracy.}

\subsection{Replay}
\label{subsec:arch_replayig}


The replaying component allows submitting recorded \textit{VM seeds} to the hypervisor. It is worth noting that the proposed framework also allows submitting crafted VM seeds, i.e., seeds built manually. When a \textit{VM seed} is submitted, the hypervisor executes the related sensitive operations according to the VM exit reason. \revision{Note} that a sequence of VM exits is usually a result of complex operations that are difficult to be reproduced at the guest level (see \sectionautorefname{}~\ref{sec:probl_statement}). 

\revision{Our idea is to reduce the replay time (to improve the efficiency) by enabling a continuous triggering of VM exits that directly stimulate the \textit{VM exit handler}, without actually executing the guest. To perform the replay, we use a \textit{dummy VM} that does not run any operation except \textit{i)} initialization of all hypervisor data structures and VMCS where the seeds will be submitted and \textit{ii)} triggering of VM exits to execute the \textit{VM exit handler} (switch from \textit{non-root mode} to \textit{root mode}). }


A possible solution to continuously submit VM exits is to let the dummy VM do a first exit and then implement a loop directly within the \textit{VM exit handler}. \revision{However, a loop in \textit{root mode} could be detected from the hypervisor as a hang, leading to forced crashes. In addition, such a direct loop avoids the VM entry (see \sectionautorefname{}~\ref{sec:background}), which occurs as the last step in the VM exit handling. The \textit{VM entry} operation includes several checks on the VMCS fields (specified in Section 26.3 in \cite{intel2009intel}) that \revision{are representative or real VM behavior and} are used to guarantee semantically-correct VM seeds submission. Thus, our replaying architecture lets the VM exit handler execute the VM entry and then forces the \textit{dummy VM} to immediately trigger another VM exit, preventing the execution of any instruction at the guest level.} 


To set up the context of the hypervisor according to the \textit{VM seed}, GPRs are rewritten in the hypervisor data structures where they are stored. The VMCS fields are rewritten ({\lmttfont VMWRITE}s) with \textit{VM seed} values if they are read ({\lmttfont VMREAD}s) again during the replay. If the VMCS fields are read-only, and cannot be rewritten, we modify only the return value of the {\lmttfont VMREAD}s.
After the \textit{VM seed} submission, the hypervisor handles sensitive operations according to the \textit{VM seed}, before executing the VM entry operation. 




\revision{Finally, we define \textit{accuracy} of the replaying mechanism as its ability to reproduce a valid VM behavior for the recorded metrics, i.e., code coverage at the hypervisor level and writes performed into the VMCS fields.} \revision{To evaluate the accuracy, IRIS allows reverting the \textit{test VM} snapshot saved at the start of recording, and using it as a starting point from which replaying \textit{VM seeds} via the \textit{dummy VM}.}



\subsection{Manager}

\revision{As already explained}, the proposed framework provides the \textit{record} and \textit{replay} \revision{operation modes}. Further, we implemented a component called \toolname{} \textit{manager}, which exposes an interface that can be used by a \textit{user-space application (CLI)} to \textit{(i)} choosing between \revision{operation modes}, i.e., \textit{record} and \textit{replay}; \textit{(ii)} retrieve \textit{VM seeds} and metrics during the \textit{record mode}; \textit{iii)} submitting \textit{VM seeds} during the \textit{replay mode}. When the \toolname{} \textit{manager} enables the \textit{record mode}, it runs a \textit{test VM} and allows it to trigger normal VM exits as they occur. 
The \textit{record mode} can be configured to store \textit{VM seeds}, metrics, or both of them. Recording can be manually or programmatically stopped after a given number of monitored VM exits. After a specific time of recording, the \toolname{} \textit{manager} allows keeping the \textit{test VM} in an idle loop, reading for a new recording session; otherwise, the \textit{test VM} continues execution with no recording enabled. In the \textit{replay mode}, \toolname{} \textit{manager} first runs a \textit{dummy VM} (optionally by reverting to a particular VM state) and then allows users to submit seeds on-demand. Also, in this case, the \textit{manager} puts the \textit{dummy VM} in an idle loop to wait for new \textit{VM seeds} to submit. When a new \textit{VM seed} is available for submission, the replaying component submits it to the hypervisor. \revision{Note that both recorded seeds and manually crafted seeds can be submitted at this step.}
Finally, the \toolname{} \textit{manager} allows enabling the \textit{replay mode} together with the \textit{record mode} enabled to store metrics while replaying. This latter is necessary to evaluate the accuracy and efficiency of recorded/crafted \textit{VM seeds} which are submitted via the \textit{replay mode}.

\section{Implementation}
\label{sec:xen_implementation}

The current version of \toolname{} is built upon the Xen hypervisor. The reference CPU architecture is Intel x86 with hardware virtualization extension VT-x. We chose Xen since it is an open-source hypervisor and supports \textit{Hardware Virtual Machine} (\textit{HVM}) mode. HVM mode tries to make full virtualization easier, using the hardware emulation to accelerate CPU virtualization (privileged instructions) and the MMU (page tables). \toolname{} is implemented as a set of patches for the Xen kernel. All \toolname{} components code is written in C language. The details of each component are briefly explained in the following. 


\subsection{Record}
\label{subsec:impl_replayig}

\revision{Currently, in IRIS we obtain code coverage at the hypervisor level via compile-time instrumentation approaches using \textit{gcov} \cite{gcov}.} The hypervisor codebase should not be instrumented as a whole since we need to avoid most sources of non-determinism, e.g., due to interrupts, kernel threads, \revision{statefulness}. We selectively instrument hypervisor components crucial for VM exit handling, such as the abstraction of vCPU, HVM domain-specific functions, and the handler of VMX-related operations. While running, the resulting instrumented binary will write its own basic block coverage to a bitmap, which is exported as a shared memory area accessible at the guest level. We remark that code coverage is cleaned up by removing hits due to the execution of our record and replay components. \revision{Further, code coverage information can be retrieved for each VM seed submitted.}
Regarding the reads and writes performed into the VMCS fields, the hypervisor uses machine instructions {\lmttfont VMREAD} and {\lmttfont VMWRITE}, wrapped respectively by Xen {\lmttfont \_vmread()} and {\lmttfont \_vmwrite()} functions. We instrument these functions by adding a callback function invocation to store pairs of VMCS \{field, value\} read or written in the shared memory area. Regarding the values of guest GPR, they are stored in Xen data structures during VM exit handling, since they are not included in the VMCS. 
\revision{The current implementation, for each VM seed, uses an array of struct to store GPR, VMCS fields read or write. The struct is defined to store: i) a flag (1 byte) that indicates the kind of data; ii) the encoding (1 byte) of GPR (15 values) or VMCS fields (147 values); iii) the value (8 bytes) for GPR or VMCS field.}
Once again, we invoke a callback function at the start of the VM exit handler execution, to also buffering this kind of data. 
The temporal metric can be retrieved using instructions to get \revision{CPU-cycles} counters. For example, Intel provides the {\lmttfont RDTSC} instruction, which reads the current value of the CPU's time-stamp counter.

\subsection{Replay}
The replaying component enables the continuous triggering of VM exits by leveraging a \textit{dummy VM} running in HVM mode. We implemented the VM exit/entry loop described in \subsectionautorefname{}~\ref{subsec:arch_replayig} by enabling the Intel \textit{VMX-preemption timer} for the \textit{dummy VM}. The \textit{VMX-preemption timer} counts down (from the value loaded by a VM entry) in VMX non-root operation, at a rate proportional to that of the timestamp counter ({\lmttfont TSC}). When the timer counts down to zero, it stops counting down and a VM exit occurs (see Sections 25.2, 25.5.1, 26.6.4 \cite{intel2009intel}). In our framework, a preemption timer value set equal to zero allows the hypervisor to preempt the \textit{dummy VM} execution before the CPU executes any instructions in the guest. 



Regarding \textit{VM seed} submission, we implement callback functions inserted at compile-time and invoked during the VM exit handling to submit GPR and VMCS values according to the \textit{VM seed} values. The GPR values are simply copied to the corresponding hypervisor data structures. The VMCS values can be written into the VMCS by invoking the {\lmttfont \_vmwrite()} function. However, this solution is adopted only for writable VMCS fields, since some of them are read-only. For the latter scenario, we instrument the function {\lmttfont \_vmread()} by inserting a callback function to replace the values returned from the readings on the VMCS with those submitted with the \textit{VM seed}.

\subsection{Manager}
The manager component consists of a \revision{backend} driver at the hypervisor level. The interface exposed to users is implemented using the \textit{hypercall} mechanism. We implemented the {\lmttfont xc\_vmcs\_fuzzing()} hypercall to trap into the hypervisor, to enable and control the recording and replaying phases. Given that, a user-space application (\toolname{} CLI) invokes such hypercall to enable the functionalities provided by \toolname{} manager. Finally, the manager uses  {\lmttfont copy\_to\_guest()} and {\lmttfont copy\_from\_guest()} Xen routines to respectively retrieve recorded VM seeds and metrics and submit \textit{VM seeds}.

\section{Evaluation}
\label{sec:evaluation}

\begin{figure*}[t]
    \centerline{\includegraphics[width=1.9\columnwidth]{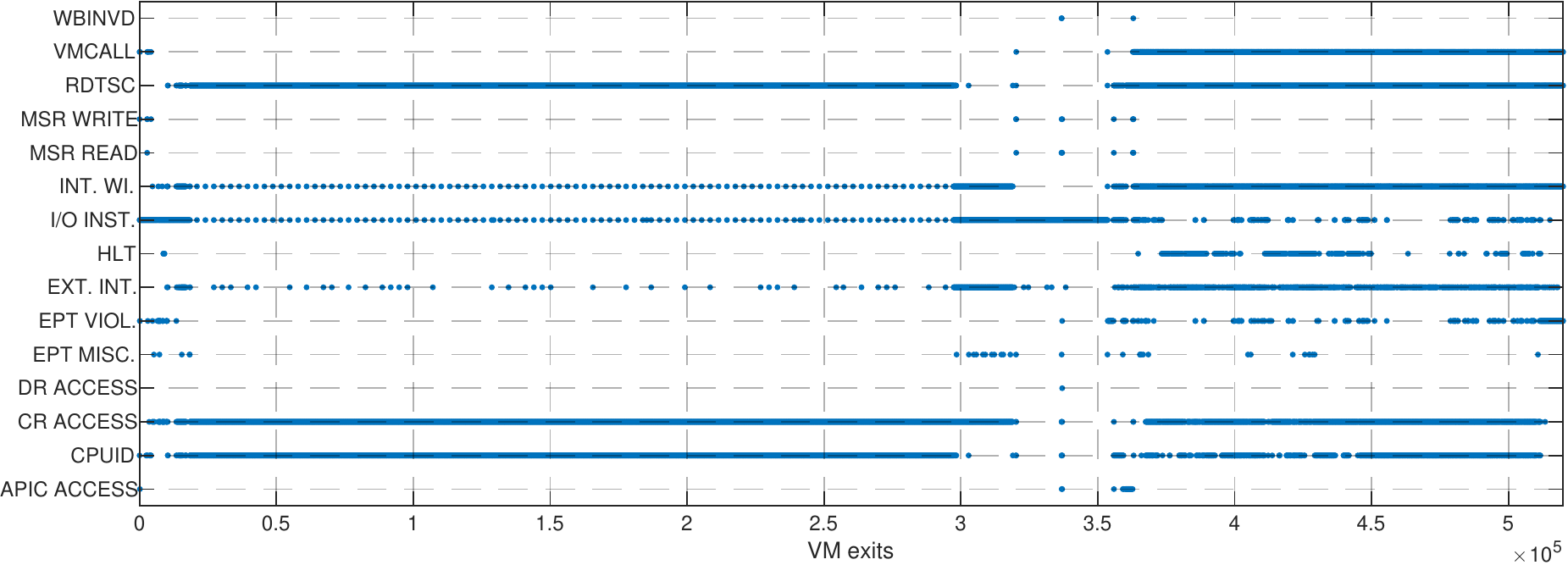}}
    \caption{VM exit reasons distribution over time during \textit{OS BOOT} workload.}
    \label{fig:distrib_vmexit_os_boot}
\end{figure*}

We perform a thorough experimental analysis of \toolname{}.
\revision{The aim is to evaluate the {accuracy} of \toolname{} at reproducing} \textit{VM behaviors} using recorded \textit{VM seeds} via the proposed replaying mechanism. To estimate the accuracy, we use the metrics provided by \toolname{} gathered during the execution of guest workloads. Further, we evaluate the efficiency in submitting recorded \textit{VM seeds} via our replaying approach in terms of the CPU time needed to execute related VM exits. Finally, we provide a \revision{proof-of-concept implementation of a fuzzer built upon \toolname{} record and replay mechanisms.}

We performed our experiments using a host machine with Intel Xeon i7-4790 @3.6Ghz, and 16GB RAM, running Linux kernel v5.10. We implemented \toolname{} on top of Xen hypervisor v4.16, running with \textit{HVM} mode in order to enable hardware-assisted virtualization. Currently, \revision{the} \toolname{} framework supports Intel VT-x hardware extensions. We run the \toolname{} \textit{manager} in \textit{Dom0}, and a guest workload to be recorded and replayed in a \textit{DomU}. This latter domain is our \textit{test VM}, while a second \textit{DomU} is mounted as the \textit{dummy VM}. Each domain \revision{(both Dom0 and DomU)} runs Linux kernel v5.10 and is set up with a single vCPU pinned on a dedicated pCPU, 1 GB RAM, and 20 GB HDD. We impose a  1-to-1 vCPU/pCPU pinning for VMs to prevent as many as possible interferences (e.g., high rate of cache misses) between the different physical cores and obtain coverage data as clean as possible.

\subsection{Workloads}
\label{subsec:workloads}

\revision{Experiments are based on workloads that have been characterized by the \toolname{} recording mechanism in terms of \textit{VM behavior}. }
\revision{For the sake of simplicity we consider a sample trace of $5000$ VM exits for each workload.} In particular, we consider \textit{i)} booting an \revision{OS kernel} (\textit{OS BOOT}); stressing \textit{ii)} CPU subsystem (\textit{CPU-bound}), \revision{\textit{iii)} memory subsystem (\textit{MEM-bound}), and \textit{iv)} I/O subsystem (\textit{I/O-bound}); \textit{v)}} keeping idle the OS (\textit{IDLE}). 

The \textit{OS BOOT} workload, specifically refers to booting the Linux kernel, and \revision{it consists of about} $520$K VM exits until the OS presents the login screen to the user. \figureautorefname{}~\ref{fig:distrib_vmexit_os_boot} details the VM exits distribution during \textit{OS BOOT} workload over time. \revision{It can be noted that \toolname{} is capable of recording all the VM exits occurring during the boot sequence. The distribution includes a sequence of VM exits (the first 10K) that are related to the BIOS emulated by Xen \cite{hvmloader}, which is not part of the OS BOOT we want to characterize. Given this, our OS BOOT trace of $5000$ VM exits starts after the last BIOS VM exit.} The \textit{CPU-bound} workload includes \revision{$5000$ VM exits triggered during the execution of CPU-intensive operations} (e.g., compute Fibonacci sequence, matrix operations, etc.). \revision{The \textit{MEM-bound} workload include $5000$ VM exits triggered during the execution of memory-intensive operations for the stack, heap, memory mapping, and shared memory, while \textit{I/O-bound} workload focuses on generic input/output.} Finally, the \textit{IDLE} workload includes $5000$ VM exits triggered during the OS idle loop. 

\begin{figure}[b!]
    \center
    \includegraphics[width=.9\columnwidth]{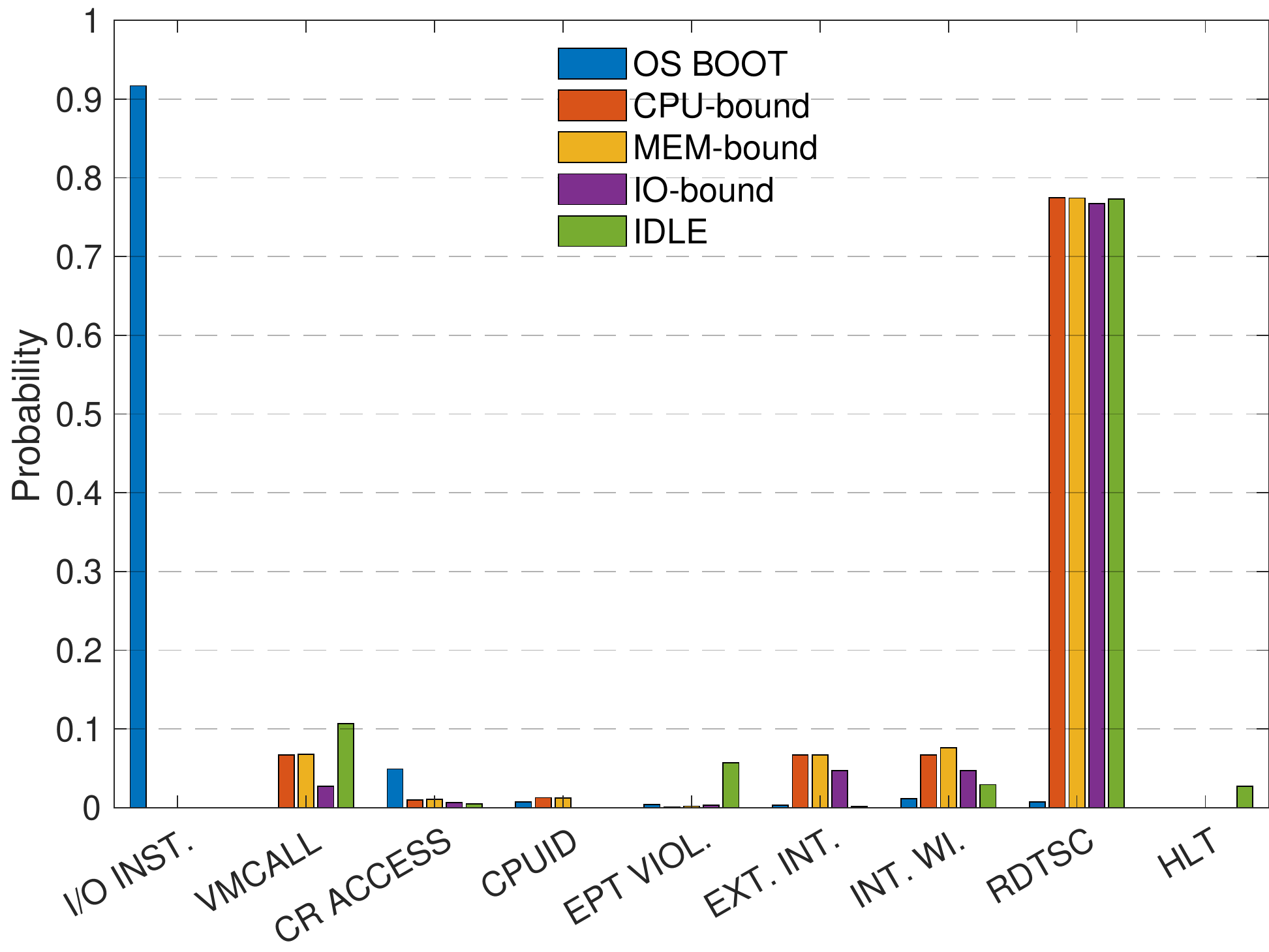}
    \caption{VM exit reasons distribution over \revision{different target workloads.}}
    \label{fig:distrib_vmexit}
\end{figure}

\figureautorefname{}~\ref{fig:distrib_vmexit} summarizes the distribution of VM exits across the target guest workloads. We can notice that the \textit{OS BOOT} workload results in VM exits that are mostly related to \textit{I/O instruction} and \textit{Control-register accesses} exit reasons. In the boot phase, the guest configures devices and the hypervisor \revision{is triggered to carry out operations for} emulation, paravirtualization, exclusive assignment, or I/O device sharing depending on the kind of virtualization approach used \cite{abramson2006intel}. In the remaining workloads (i.e., \textit{CPU-bound} \revision{, \textit{MEM-bound}, \textit{I/O-bound},} and \textit{IDLE}), almost $80\%$ of VM exits are related to {\lmttfont RDTSC} instructions, which are related to kernel operations needed for timekeeping and implementing scheduling routines \cite{linux_timekeeping}. Further, the \textit{IDLE} workload is characterized by some {\lmttfont HLT} VM exits, basically due to the implementation of the \textit{idle loop} in the Linux kernel.

\revision{\figureautorefname{}~\ref{fig:distrib_vmexit} also reveals that the hypervisor intervention is relegated to few critical VM exits, which are common across different and heterogeneous workloads. For the sake of representativeness, we further analyzed benchmarks provided in \cite{wei2019hyperbench}, which exhibit similar VM exits distributions obtained above. For simplicity, the analysis provided in the next sections targets only the \textit{CPU-bound} workload since it has several commonalities with the MEM-bound and I/O-bound experimented workloads.}

\subsection{\revision{Accuracy}}
\label{subsec:effect}

\revision{We analyze how accurate is \toolname{} in the automatic recording and replaying of \textit{VM behaviors} using} key metrics mentioned in \sectionautorefname{}~\ref{sec:architecture}, i.e., code coverage and the pair \{VMCS field, value\} written. For this purpose, we aim to reveal the difference between metrics gathered both in the recording and replaying phases \revision{for each VM seed obtained during the execution of target workloads described in \sectionautorefname{}~\ref{subsec:workloads}. Note that we use the same VM snapshot as the starting state for the record and replay phases to unbias the accuracy evaluation}.


\begin{figure*}[t!]

    \centering
     \includegraphics[width=.8\columnwidth]{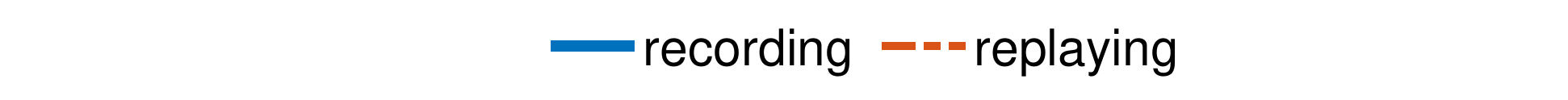}
     
     \centering
     \begin{subfigure}[b]{0.329\textwidth}
         \centering
         \includegraphics[width=\linewidth]{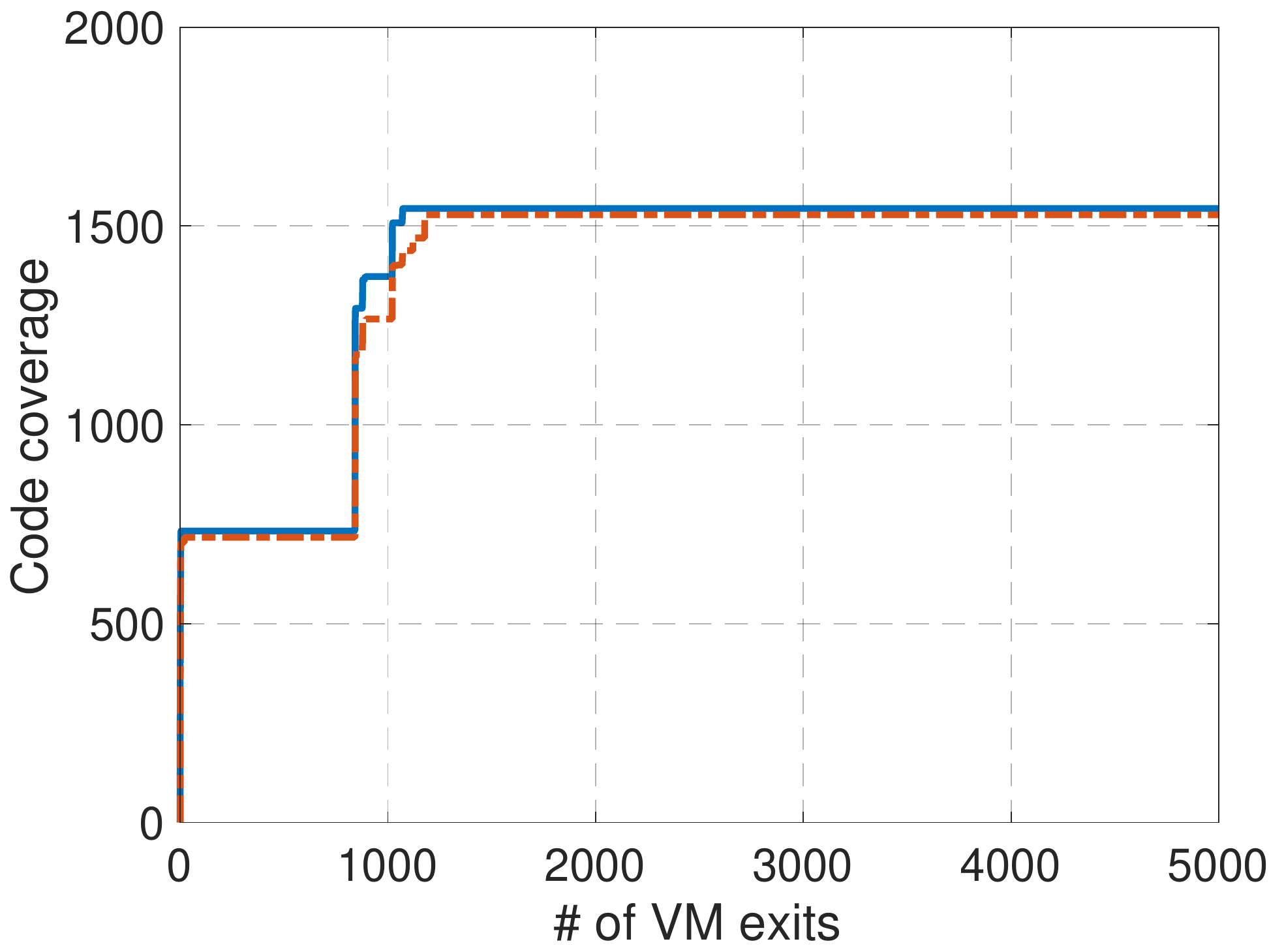}
         \caption{OS BOOT}\label{fig:boot_eff}
     \end{subfigure}
     \hfill
     \begin{subfigure}[b]{0.329\textwidth}
         \centering
         \includegraphics[width=\linewidth]{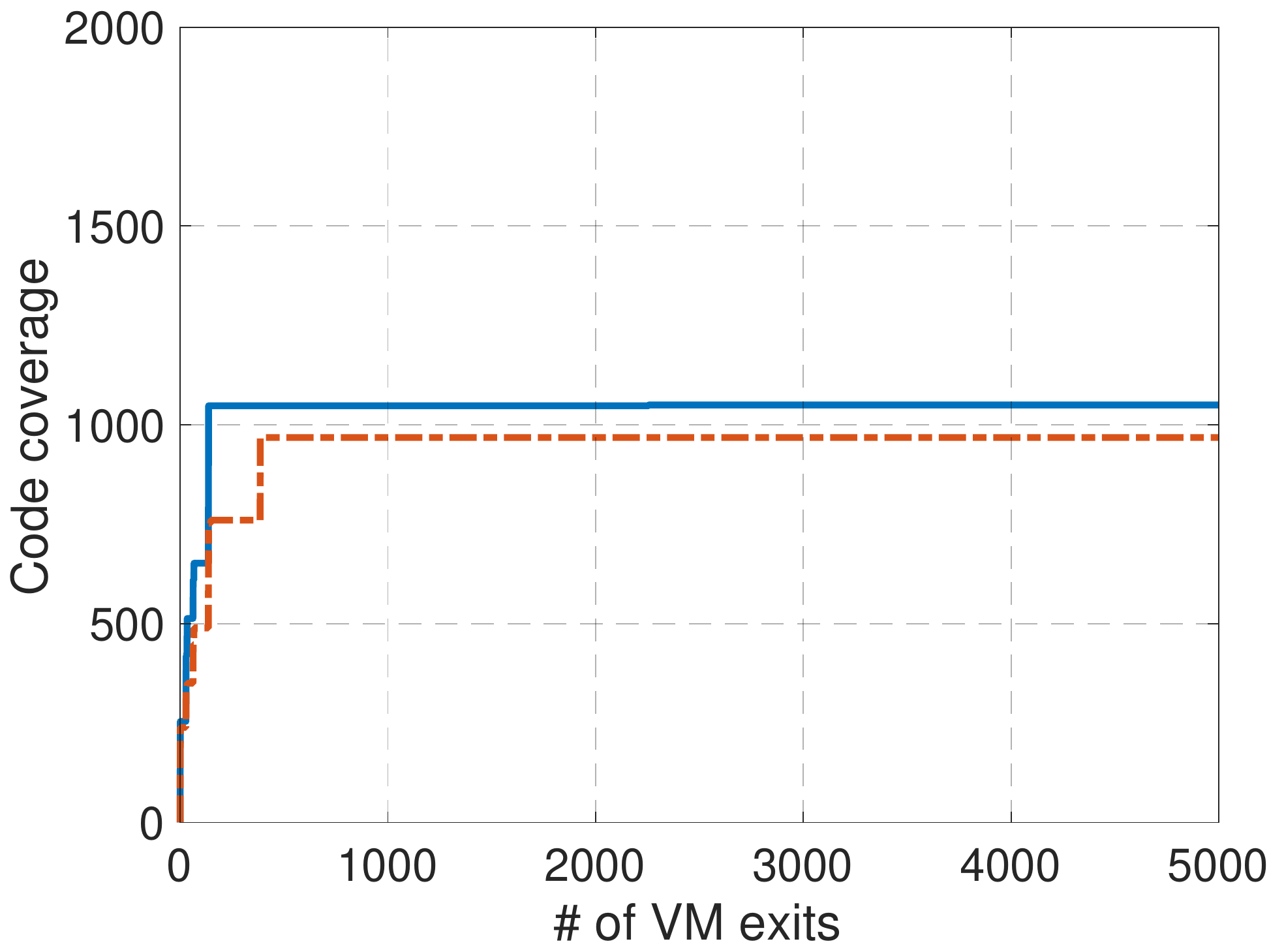}
         \caption{CPU-bound}\label{fig:cpu_eff}
     \end{subfigure}
     \hfill
     \begin{subfigure}[b]{0.329\textwidth}
         \centering
         \includegraphics[width=\linewidth]{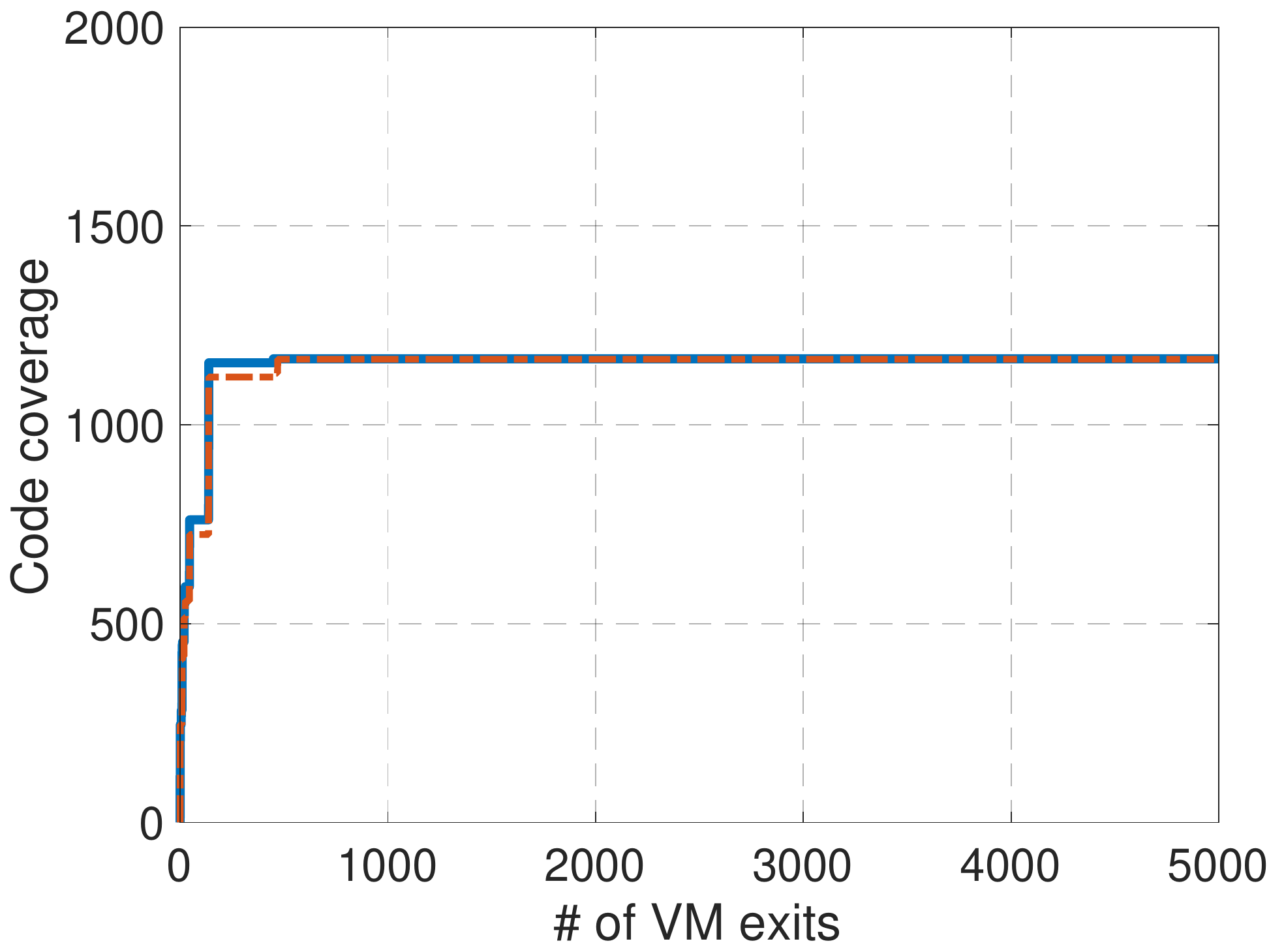}
         \caption{IDLE}\label{fig:idle_eff}
     \end{subfigure}
        \caption{Cumulative code coverage across \textit{OS BOOT}, \textit{CPU-bound}, and \textit{IDLE} workloads.}
        \label{fig:effectiveness_workload}
\end{figure*}

\revision{Regarding accuracy in terms of code coverage,}
\figureautorefname{}~\ref{fig:effectiveness_workload} shows the results across the target workloads. The recording curve identifies the cumulative coverage trend for \revision{\textit{VM seeds}} recorded, in which we evaluate the unique lines of code discovered during VM exit handling, for each \textit{VM seed}. Instead, the replaying curve identifies the cumulative coverage trend \revision{for the replaying of the same VM seeds}. The code coverage fitting at the end of replaying \revision{\textit{VM seeds}} is \revision{\textit{$99.9\%$, $92.1\%$, and $98.9\%$}} for \textit{OS BOOT}, \textit{CPU-bound}, and \textit{IDLE} workloads, respectively. 

\begin{figure}[!b]
    \centering
    \includegraphics[width=.9\columnwidth]{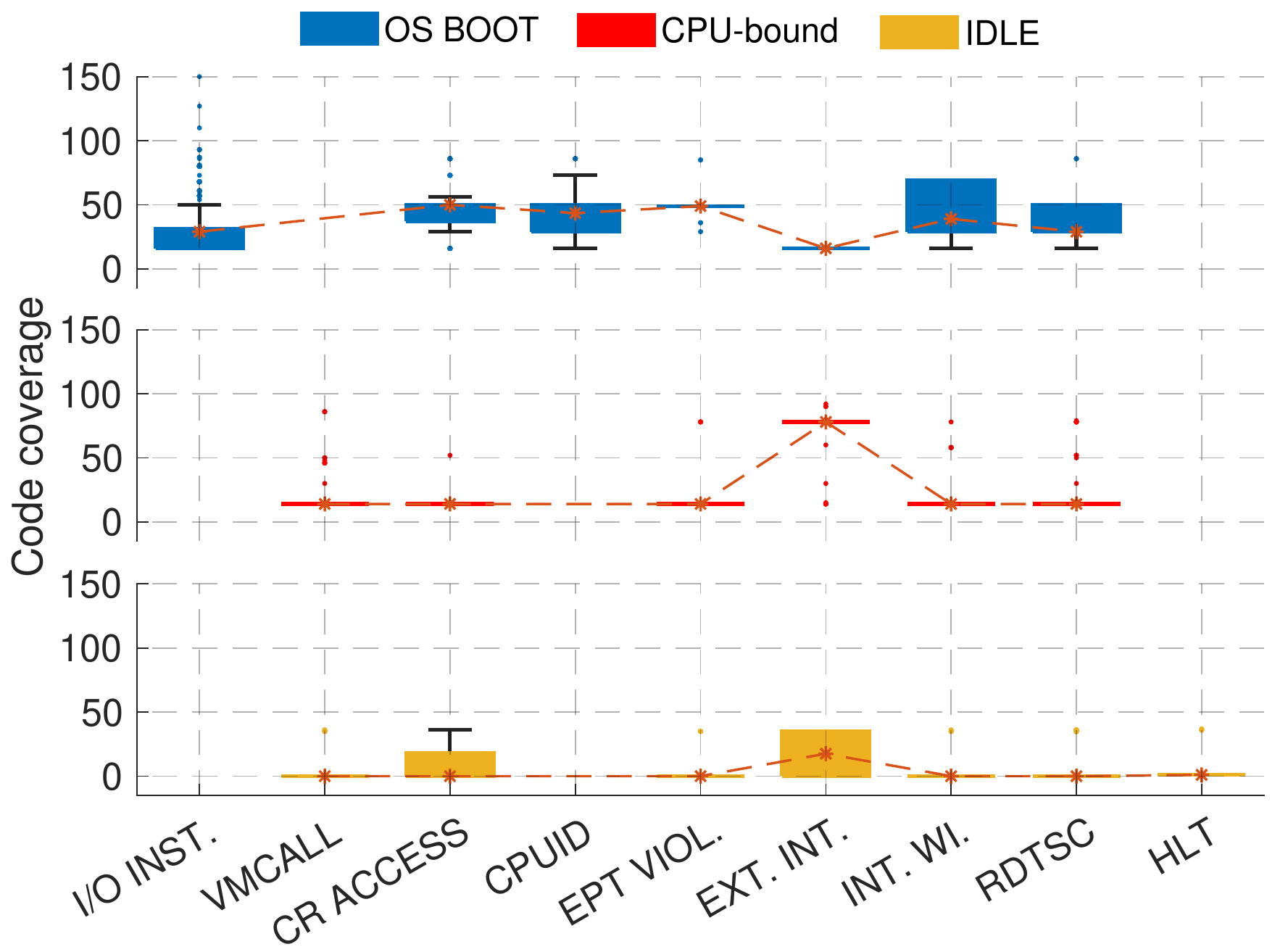}
    \caption{Code coverage differences by VM exit reason across targeted workloads.}
    \label{fig:missing_distrib_vmexit}
\end{figure}

\revision{Despite we achieved high accuracy in code coverage, we further analyze the remaining differences qualitatively. 
 \figureautorefname{}~\ref{fig:missing_distrib_vmexit} shows the code coverage differences across the three target workloads, which we clustered by VM exit reasons. The code coverage data may be affected by sources of non-determinism due to asynchronous events that interrupt the hypervisor (in VMX root mode) during the VM exits handling. The minimum code coverage difference ranges from $1$ to $30$ LOC. By analyzing such cases, we point out that such differences are related to the \textit{local Advanced Programmable Interrupt Controller (\textit{"vlapic.c"})}, \textit{interrupt handling} (\textit{"irq.c"}), and \textit{virtual timer} ({\lmttfont vpt.c}) Xen components. We can treat such coverage differences as noise to filter out.}

\revision{We also investigate the cases when the code coverage differences are greater than 30 LOC. The frequency of these cases (filtering the repeated VM seeds in a workload) is $0.36\%$, $0.18\%$, and $1.16\%$ for \textit{OS BOOT}, \textit{CPU-bound}, \textit{IDLE} workloads, respectively. 
These differences refer to the \textit{HVM instruction emulator} ({\lmttfont "emulate.c"}) and \textit{VM exit handler} ({\lmttfont "intr.c"}, and {\lmttfont "vmx.c"}) Xen components.}
\revision{This behavior can be due to the recorded VM seeds that are linked to memory-related VM exits. For example, VMCS fields like \textit{Global} and \textit{Local Descriptor Table Registers} (GDTR and LDTR) include references to the memory of "exited" guest VM. Such values can be dereferenced by the hypervisor during exit handling.}

\revision{We further evaluate the IRIS accuracy, by focusing on the {\lmttfont VMWRITE}s metric, which provides a more fine-grained measure of actual VM state changes (guest-state area) from the point of view of the VMCS and VMX operations.}

\begin{figure}[!b]
    \centerline{\includegraphics[width=.85\columnwidth]{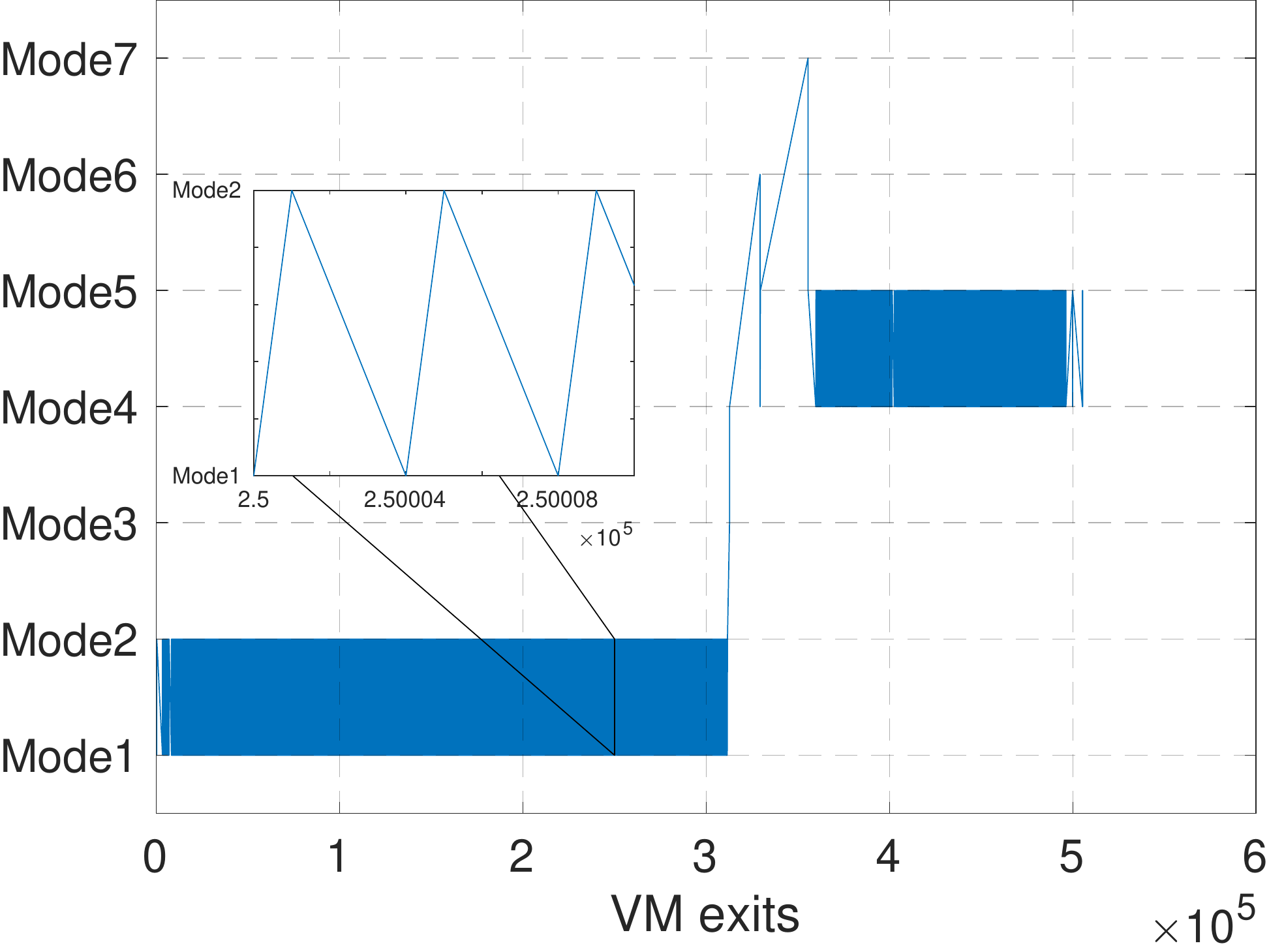}}
    \caption{Operating modes and virtual CPU states across VM exits during \textit{OS BOOT} workload.}
    \label{fig:vmwrites_fitting_on_cr0}
\end{figure}

Focusing on the entire \textit{OS BOOT} workload (see \figureautorefname{}~\ref{fig:distrib_vmexit_os_boot}), the OS switches operating modes and CPU states several times. In that case, the fitting on the executed {\lmttfont VMWRITE}s on the VMCS guest-state area is $100\%$. Indeed, \figureautorefname{}~\ref{fig:vmwrites_fitting_on_cr0} shows an example for {\lmttfont VMWRITE} operations both recorded and replayed against the control register zero (i.e., {\lmttfont CR0}). Each of the modes represents a set of states held by the CR0 register.

\begin{figure*}[t]
     \centering
     \includegraphics[width=.3\columnwidth]{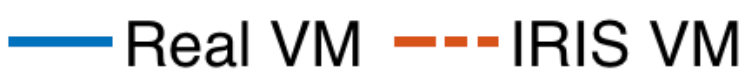} 
     
     \begin{subfigure}[b]{0.329\textwidth}
         \centering
         \includegraphics[width=\linewidth]{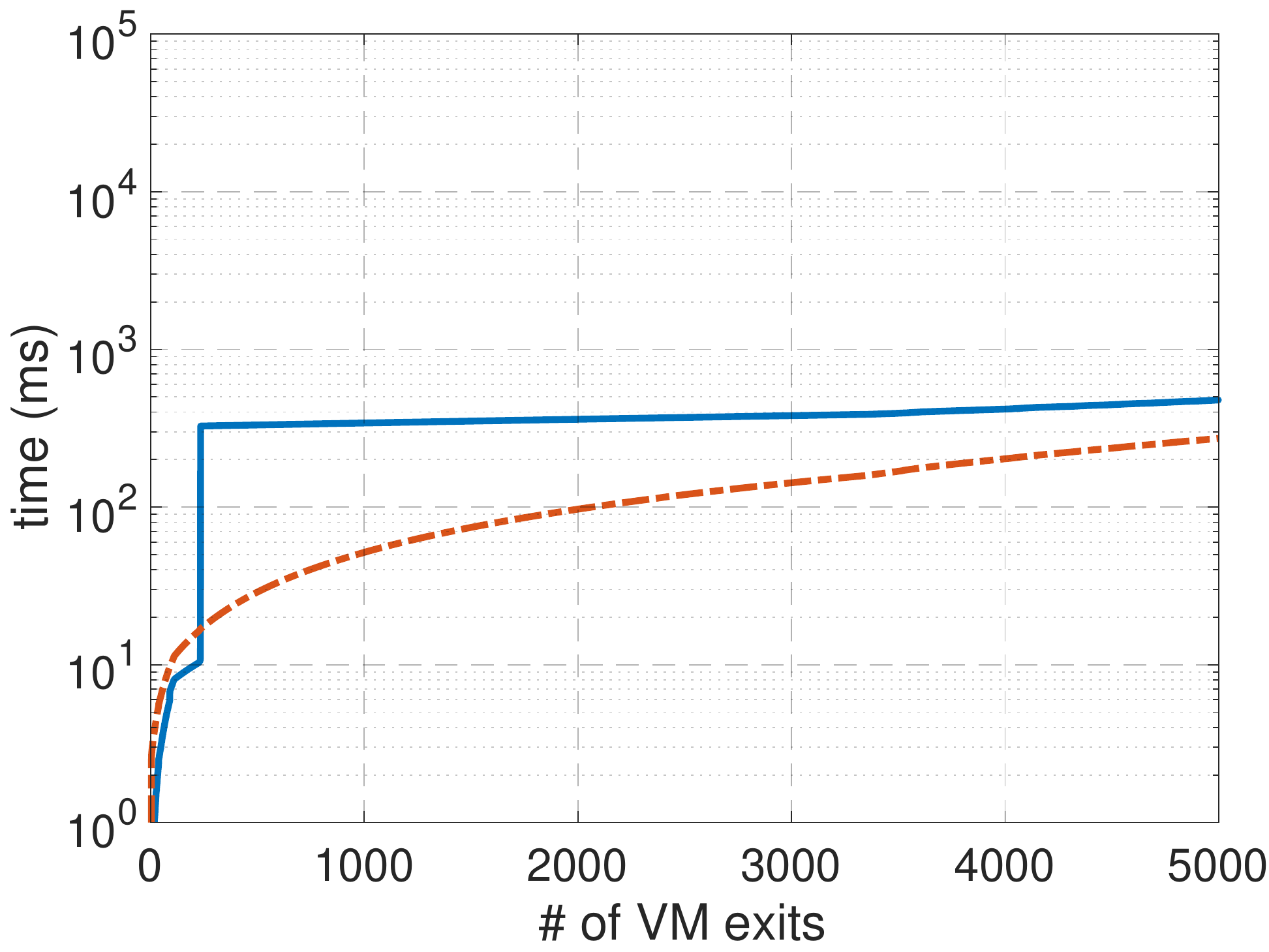}
         \caption{OS BOOT}\label{fig:efficency_boot}
     \end{subfigure}
     \hfill
     \begin{subfigure}[b]{0.329\textwidth}
         \centering
         \includegraphics[width=\linewidth]{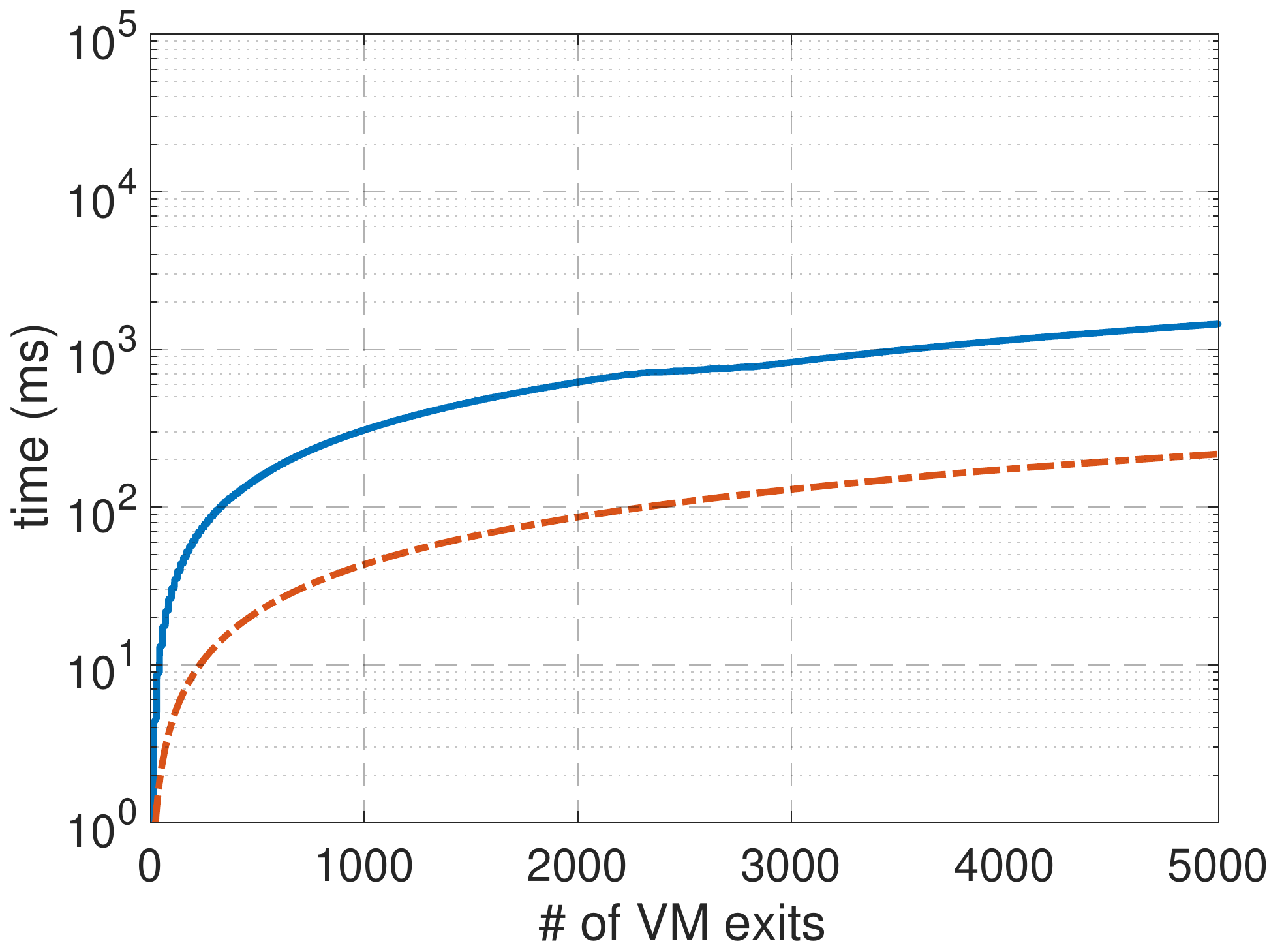}
         \caption{CPU-bound}\label{fig:efficency_cpu}
     \end{subfigure}
     \hfill
     \begin{subfigure}[b]{0.329\textwidth}
         \centering
         \includegraphics[width=\linewidth]{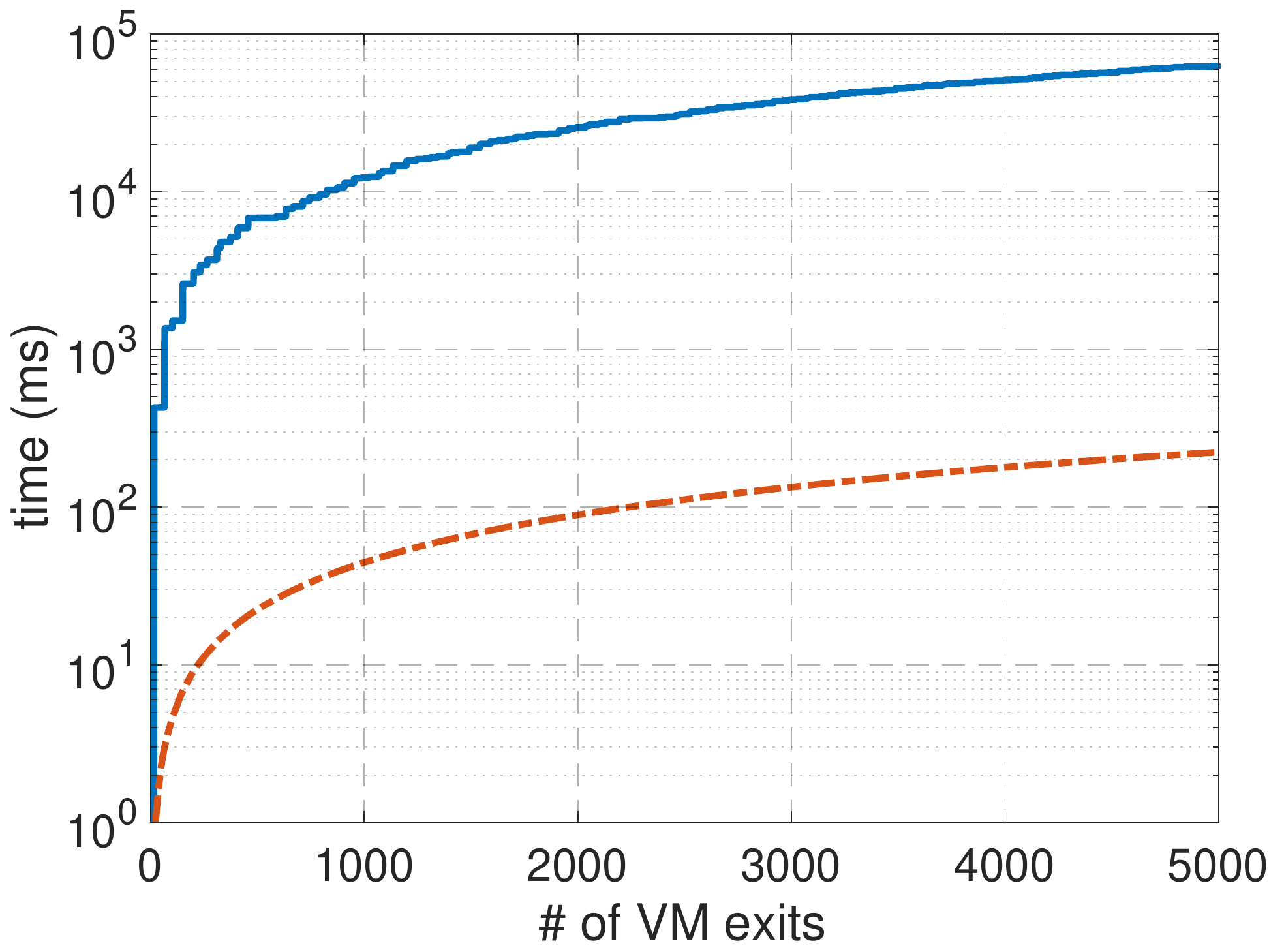}
         \caption{IDLE}\label{fig:efficency_idle}
     \end{subfigure}
        \caption{Performance in submitting \textit{VM seeds} across \textit{OS BOOT}, \textit{CPU-bound}, and \textit{IDLE} workloads.}
        \label{fig:efficency_workloads}
\end{figure*}

In particular, \textit{Mode1} and \textit{Mode2} indicate \textit{real mode} and \textit{protected mode}, respectively. \textit{Mode3} specifies \textit{protected mode} with \textit{paging enabled}, \textit{Mode4} includes \textit{Mode3} with alignment checking performed, \textit{Mode5} includes \textit{Mode4} with test of task switch flag, \textit{Mode6} includes \textit{Mode4} and caching enabled, \textit{Mode7} includes \textit{Mode5} and caching disabled. Results suggest that the proposed recording approach allows us to generate seeds that closely follow real \textit{VM behaviors} of guest execution. 
\revision{Finally, we run an experiment to provide evidence that replaying recorded VM seeds allows reaching the same hypervisor state as in the real guest execution. We replay \textit{CPU-bound} and \textit{IDLE} workloads from a \textit{i)} VM state without booting the OS, and from a \textit{ii)} VM state reached by replaying the recorded \textit{OS BOOT} VM seeds. In the former case, the \textit{dummy VM} crashes (Xen logs: {\lmttfont bad RIP for mode 0}, where {\lmttfont mode 0} is \textit{Mode1} in \figureautorefname{}~\ref{fig:vmwrites_fitting_on_cr0}), while in the latter case both the \textit{CPU-bound} and \textit{IDLE} workloads complete.}

\subsection{\revision{Efficiency}}
\label{subsec:efficiency}

\revision{Seeds submission is fundamental in developing fuzzers since it heavily impacts fuzzing efficiency \cite{zhu2022fuzzing}. To this end, we estimate how \textit{IRIS} is efficient in reaching VM states compared to the real guest VM execution.} We performed this analysis across workloads described in \subsectionautorefname{}~\ref{subsec:workloads}, by running the experiments $15$ times for statistical significance purposes, obtaining the same results with a high level of confidence (p-value $< 0.05$).

\figureautorefname{}~\ref{fig:efficency_workloads} shows the time needed to \revision{submit VM seeds by real guest VM execution (see \textit{Real VM} in \figureautorefname{}~\ref{fig:efficency_workloads} and by using the IRIS replaying mechanism (see \textit{IRIS VM} in \figureautorefname{}~\ref{fig:efficency_workloads})}. \revision{The results} show that our replaying mechanism can replay real guest VM behaviors efficiently, with a \revision{percentage decrease of $42.5\%$ ($0.27s$ vs $0.47s$), $85.4\%$
($0.21s$ vs $1.44s$), and $99.6\%$ ($0.22s$ vs $62.61s$)}, for \textit{OS BOOT}, \textit{CPU-bound}, and \textit{IDLE} workloads respectively. \revision{It is worth noting that the \textit{OS BOOT} exhibits the main differences} in the first $1000$ VM exits, in which the kernel OS spends time running guest operations that do not require the hypervisor intervention. These non-sensitive instructions delay substantially the subsequent VM exits that eventually need to be handled by the hypervisor; thus, there is a non-negligible latency in discovering the same coverage compared to the replayed workload. In general, the throughput of our replaying mechanism is roughly linear, as also confirmed by \revision{the time distributions} to replay \textit{CPU-bound}, and \textit{IDLE} workloads (see \figureautorefname{}~\ref{fig:efficency_cpu} and \figureautorefname{}~\ref{fig:efficency_idle}). \revision{These workloads require less hypervisor intervention (VM exits handling)}, thus the IRIS replaying is even better to discover the same coverage as real guest execution in less time, with a speedup factor of $6.8\times$ and $294\times$ for \textit{CPU-bound} and \textit{IDLE} workloads, respectively. 

\begin{figure}[!b]
    \center
    \includegraphics[width=.8\columnwidth]{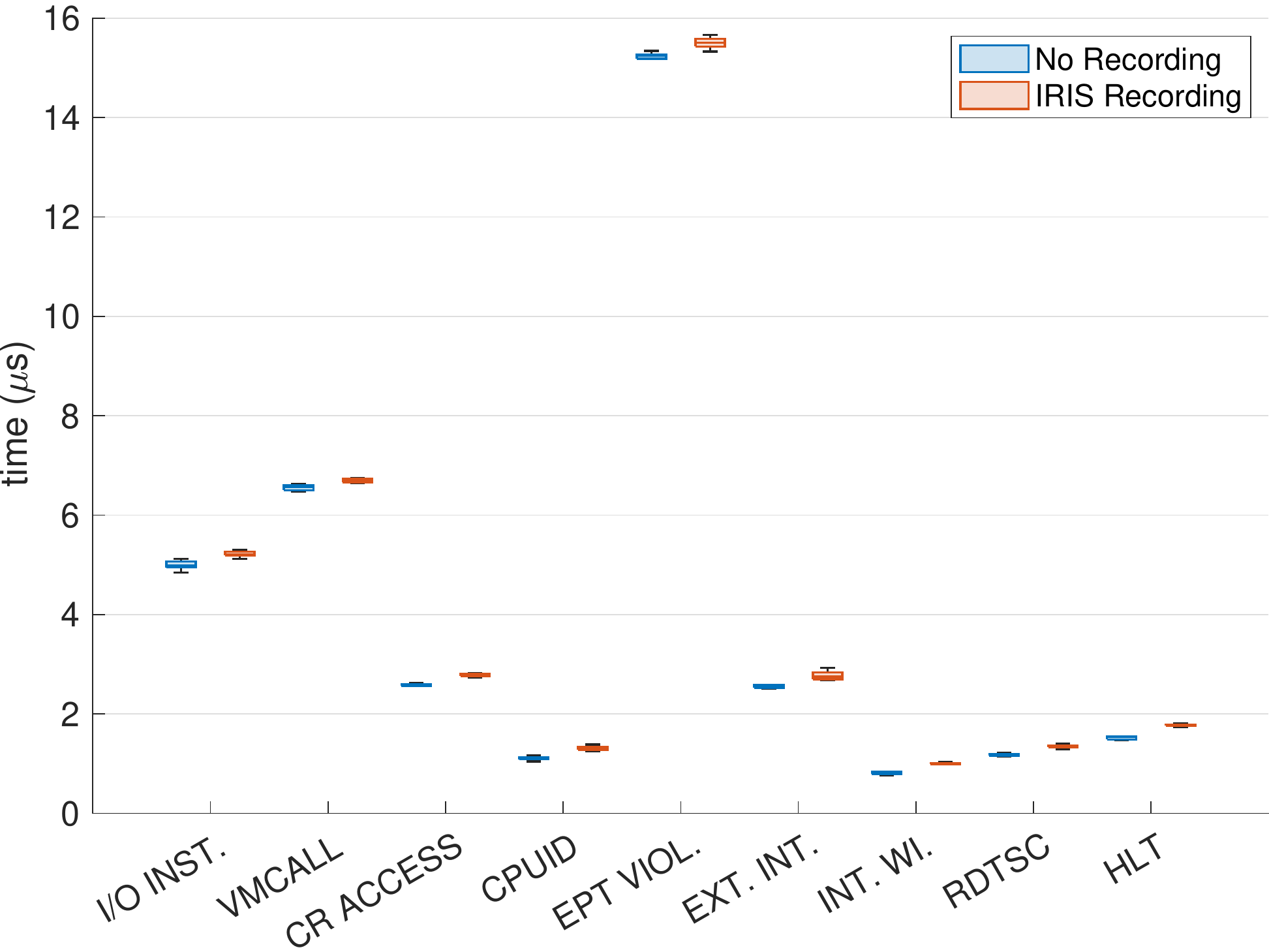}
    \caption{\revision{The temporal overhead, for each \textit{VM exit}, induced by IRIS recording.}}
    \label{fig:overhead_rec}
\end{figure}

In addition, we also measure an \textit{ideal replaying throughput} to have an upper bound for estimating the maximum replaying efficiency. We computed \revision{this value by} running the preemption timer \revision{VM} exits in the same number \revision{of the VM exits needed per workload (i.e., $5000$ VM exits)}, and measure the time needed to handle them. We obtained $0.1s$ ($\sim350M$ CPU cycles for our testbed), which means $50K$ VM exits/$s$. Comparing to this \textit{ideal replaying throughput}, we \revision{obtained} a percentage difference of $63\%$ (\revision{$18.518$ VM exit/$s$}), $52\%$ (\revision{$23.809$ VM exits/$s$}), and $55\%$ (\revision{$22.727$ VM exits/$s$}) for \textit{OS BOOT}, \textit{CPU-bound}, and \textit{IDLE} workloads, respectively. However, \revision{such difference} does not contemplate the logic behind replaying mechanisms, but it is useful to make new \revision{room for improvements}.



\subsection{\revision{Performance Overhead}}

\revision{About the overhead induced by the IRIS recording process, we first analyze the target workloads (i.e., \textit{OS BOOT}, \textit{CPU-bound}, and \textit{IDLE}) with the aim to reveal the temporal overhead for each VM exit. We run the workloads $10$ times, taking the median values of time needed by the Xen \textit{VM exit handler} to serve a specific VM exit. \figureautorefname{}~\ref{fig:overhead_rec} shows the boxplots across the VM exits handled during workload execution, with and without IRIS recording activated. The results show a very small overhead, ranging from $1,02\%$ to $1,25\%$ percentage increases in the best and worst cases, respectively.
Concerning the IRIS memory overhead induced during recording/replaying, we need to consider the size of the \textit{VM seed} for each VM exit and the reads/writes performed on the VMCS. In the worst case, we experimented $32$ VMREAD/VMWRITE operations on the VMCS across all the target workloads, obtaining a VM seed size of $470$ bytes for each VM exit. The current implementation of the IRIS recording pre-allocates a heap memory equal to the VM seed size in the worst case (i.e., 470 bytes) for each VM exit to be recorded. Instead, the IRIS replaying allocates exactly the needed heap memory for each VM seed recorded since we know in advance the number of VMREAD/VMWRITE operations performed on the VMCS.}

\section{IRIS-based Fuzzer Prototype}
\label{subsec:eval_fuzzer}



\revision{We build a proof-of-concept (PoC) to show the potential of using \toolname{} to effortlessly run fuzzing experiments on Xen, as an example of a hardware-assisted solution. The aim is to show that the PoC fuzzer can discover new code coverage and detect anomalous hypervisor behaviors.
The fuzzing logic includes \textit{i)} adopting the IRIS replay mechanism to move into valid VM states by utilizing \textit{VM seeds} obtained during the recording of target workloads (i.e., \textit{OS BOOT}, \textit{CPU-bound} and \textit{IDLE}), and ii) mutating a specific VM seed by corrupting VMCS fields and GPR.}
\revision{According to the hypervisor fuzzing literature, we consider the guest VM untrusted. Specifically, the guest VM operations affect directly the VMCS (guest state) and indirectly the hypervisor control flow.}

\subsubsection{\revision{Test cases}}
\begin{figure}[t]
    \center
    \includegraphics[width=\columnwidth]{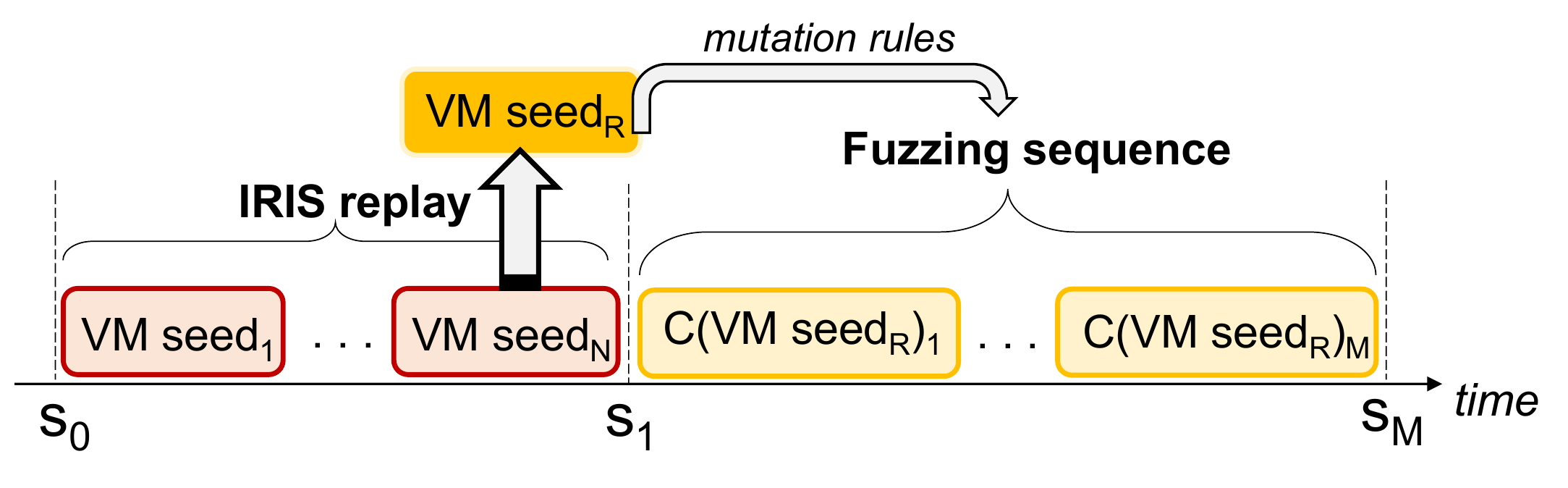}
    \caption{\revision{Test cases structure in the IRIS-based fuzzer prototype.}}
    \label{fig:testcase}
\end{figure}

\revision{The test cases we plan are characterized by the following factors: \textit{i)} the replayed \textit{VM behavior} $W$ of target workloads, a target \textit{VM seed} $VM seed_{R}$ took randomly within the \textit{VM behavior}, and the \textit{VM seed} area $A = \{VMCS, GPR\}$, of $VM seed_{R}$, to mutate.
Each test case starts from an initial VM state $s_0$ of $W$ (i.e., by starting the VM). \revision{Next, the fuzzing logic uses IRIS to replay the \textit{VM behavior} until $VM seed_{R}$ is reached, to move to the linked VM state (see $s_1$ in \figureautorefname{}~\ref{fig:testcase}).} At this point, the fuzzer mutates the $VM seed_{R}$ by generating $M$ (set equal to $10000$) mutated versions, defining the \textit{fuzzing sequence} as $C(VM seed_{R})_{1}, ..., C(VM seed_{R})_{M}$, which moves the hypervisor into an unseen state $s_M$. Such a sequence can be submitted via the IRIS replay mechanism, according to chosen \textit{mutation rules}, as depicted in \figureautorefname{}~\ref{fig:testcase}.}

\subsubsection{\revision{Mutation rules}}
\revision{The structure of test cases described above includes the \textit{fuzzing sequence} to be submitted. These mutations focus on a specific \textit{VM seed} area (i.e., VMCS or GPR) of the $VM seed_{R}$. 
The mutation rule we adopt includes a \textit{single bit-flip} in \textit{VM seed} area. Specifically, the fuzzer randomly selects a VMCS field or a general-purpose register and then bit-flip the value (e.g., {\lmttfont 0xFFFFFFF0} to {\lmttfont 0xFFFFFFF1}).} 

\subsubsection{\revision{Failure modes}}
\revision{By using scripts that analyze hypervisor behavior and logs, the PoC fuzzer can detect failures occurring during the execution of test cases, that we classify as hypervisor or VM crashes. These can be due to double faults, an invalid operation, page faults, etc.
In these cases, the test case, as well as the submitted VM seeds, are saved for further investigation with the aim of crash analysis to reveal potential bugs in the source code.}

\subsubsection{\revision{Results}}
\revision{\tableautorefname{}~\ref{tab:fuzzer_exp} shows the new code coverage discovered by running planned test cases, as explained earlier. 
The code coverage we consider as the baseline is discovered by the single $VM seed_{R}$, while each cell (i.e., a test case) of \tableautorefname{}~\ref{tab:fuzzer_exp} shows the percentage increase of code coverage discovered by submitting the \textit{fuzzing sequence}. In all tests, we can observe newly discovered coverage, with a significant increase in the \textit{OS BOOT} case, due to the complexity of the workload itself.  Note that code coverage information can be retrieved for each VM seed submitted.
Regarding failures, we observed VM or hypervisor crashes in respectively $1\%$ and $15\%$ of the tests when the VMCS is mutated. A small number of VM crashes has also been observed when mutating the GPR together with a {\lmttfont CR ACCESS} (as exit reason). In all other cases, the hypervisor is not affected by the mutation. 
The results show how the IRIS-based fuzzer PoC can discover new hypervisor code coverage and crashes, by planning a few test cases with a naive mutation rule. The manual effort in building fuzzing seeds is negligible since we obtained them by leveraging the IRIS recording mechanism. Furthermore, seed submission is done by reusing the IRIS replay mechanism with no other external tools.
}

\begin{table}[!t]
\color{black}
\caption{New code coverage discovered across test cases by using IRIS-based fuzzer prototype.}
\label{tab:fuzzer_exp}
\begin{center}
\begin{tabular}{ |c| c| c| c| c|c|c| }
\hline 
\textbf{} Exit & \multicolumn{2}{|c|}{\textbf{OS BOOT}} & \multicolumn{2}{|c|}{\textbf{CPU-bound}} & \multicolumn{2}{|c|}{\textbf{IDLE}} 
\\
 Reason & VMCS & GPR & VMCS & GPR & VMCS & GPR
\tabularnewline
\hline
 \textbf{EXT. INT.} & +122\% &+76\%  &+7\% &+3\% & +7\% & +3\%
\tabularnewline
\hline
 \textbf{INT.WI.} & +115\% &+61\%  &+6\% & +3\% &+6\% & +3\%
\tabularnewline
\hline
\textbf{CPUID.} &+124\% &+71\%   &+14\% &+2\%  & -&  -
\tabularnewline
\hline
 \textbf{HLT} & - & -  &-  & - &+7\%  &+2\%
\tabularnewline
\hline
 \textbf{RDTSC} &+120\% &+69\%  &+17\% &+2\% &+17\% &+2\%
\tabularnewline
\hline
 \textbf{VMCALL} &- &- &+18\% & +3\% &+16\% &+3\%
 \tabularnewline
\hline
 \textbf{CR ACC.} &+10\% &+63\% & +13\% &+2\%  &+10\% &+2\%
\tabularnewline
\hline
 \textbf{I/O INST.} & +8\% &+58\%  &- & - & -&  -
\tabularnewline
\hline
 \textbf{EPT VIOL.}&+22\% &+50\%  &+13\% &+1\% &+18\% &+10\%

\tabularnewline
\hline
\end{tabular}
\end{center}
\end{table}

\newcommand{\dcrash}[1]{{\textcolor{yellow} {#1}}}
\newcommand{\bugon}[1]{{\textcolor{red} {#1}}}
\newcommand{\no}[1]{{\textcolor{green} {#1}}}

\section{Related work}
\label{sec:related}

At the best of our knowledge, \toolname{} is the first framework that enables the record and replay in hardware-assisted virtualization. Its main difference from previous studies is that it does not build the seed manually. Instead, it allows recording the VM executions to learn valid VM seeds, and it allows following the entire hardware-assisted virtualization behaviors. In this section, we discuss previous studies on hypervisor testing and record and replay approaches for security applications.

\subsection{CPU Virtualization testing}

\textit{Amit et al.} ~\cite{Amit2015VirtualCV} propose to apply the testing environment of CPU vendors to hypervisors, however, they need an intimate awareness of x86 architecture to generate comprehensive test cases. \textit{PokeEMU} \cite{Yan2018FastPS} generates CPU test cases for virtual CPU implementations applying symbolic execution exclusively to an executable specification, without considering the implementation. However, its main targets \revision{are hypervisor with no hardware-assisted virtualization}. Similarly, \textit{MultiNyx} \cite{Fonseca2018MultiNyxAM} generates test cases \revision{focusing on hardware-assisted virtualization, by applying dynamic symbolic execution.} However, \textit{MultiNyx} records multiple traces between VM and VMM context incurring a \revision{high performance overhead}. \textit{HyperFuzzer} \cite{Ge2021HyperFuzzerAE} is a hybrid fuzzer for virtual CPUs. Both \textit{HyperFuzzer} and \textit{MultiNyx} are snapshot-based fuzzer. Its main difference from \cite{Fonseca2018MultiNyxAM, Yan2018FastPS} is that it avoids the overhead of a full hypervisor execution track, relying on instrumentation. Instead, it only records the program’s control flow by using commodity hardware tracing. \revision{These studies construct the initial fuzzing seeds manually based on expert knowledge. In addition, they do not focus on I/O device virtualization behaviors.}

\subsection{Device Virtualization testing}

The following studies do not mutate the VM’s architectural state. This can limit their testing coverage, as the hypervisor depends on the VM’s architectural state when emulating an operation. \textit{Schumilio et al.} \cite{Schumilo2020HYPERCUBEHH} first discover the available hypervisor interfaces via a custom OS, then they test such interfaces through a black-box fuzzer based on a custom bytecode interpreter which accelerates the input generation phase. Once again, the fuzzing seeds are built manually. \textit{Nyx} \cite{Schumilo2021NyxGH} tests the hypervisor target via nested virtualization using KVM. In addition, \textit{Nyx} uses grammar rules to specify the structure of the target emulated devices. \revision{Relying on manual input grammars per device requires manual work to specify grammar rules \cite{Myung2022MundoFuzzHF}}, thereby several studies record the interactions between the guest operating system and the device \cite{Myung2022MundoFuzzHF, Henderson2017VDFTE, Pan2021VShuttleSA, Bulekov2022MorphuzzB}. \textit{Henderson et al.} \cite{Henderson2017VDFTE} selectively instrument the code of a given virtual device, and perform a record and replay of the only memory-mapped I/O (MMIO) activity of the virtual device in QEMU. \textit{VShutlle}, \textit{Morphuzz}, and \textit{MundoFuzz} \cite{Pan2021VShuttleSA, Bulekov2022MorphuzzB, Myung2022MundoFuzzHF} fuzz the entire emulated device input interface including DMA interactions. Contrary to MMIO and PIO interactions that call the hypervisor intervention interrupting the VM (VM exit), the DMA does not interrupt the VM. Indeed, both the work \cite{Pan2021VShuttleSA, Bulekov2022MorphuzzB} instrument the DMA API of the Hypervisor to target the dynamic memory regions where the DMA is working. Instead, \textit{MundoFuzz} \cite{Myung2022MundoFuzzHF} collects IO instructions and DMA operations within the guest operating system without hypervisor instrumentation. MundoFuzz \cite{Myung2022MundoFuzzHF} fuzz the hypervisor with grammar-awareness using automatic grammar inference. Hypervisor grammars have hidden input semantics, and \textit{MundoFuzz} finds the causal relationships between the inputs through experiments (statistical learning). Additionally, the recorded inputs could be interleaved from asynchronous events (e.g., the timer interrupts) that generate coverage noises. \textit{MundoFuzz} deletes this noise through differential learning.

\subsection{Record and replay}
In fuzzing, the record and replay is an effective way to learn the grammar of the target system \cite{Myung2022MundoFuzzHF, Henderson2017VDFTE, Pan2021VShuttleSA, Shen2022DrifuzzHB}.
However, record and replay are also adopted in security to analyze and debug execution traces. Record and deterministic Replay (RnR) is a popular architectural technique \cite{Bressoud1996HypervisorbasedFT, Dunlap2002ReVirtEI, Chow2008VMwareDecouplingDP, Dunlap2008ExecutionRO, Shalabi2018RecordReplayAA, Wang2022ClusterRRAR}. The RnR injects the recorded events at the correct times, enforcing a deterministic execution (Replay). RnR is used for several reasons. For instance, when the system adopts no precise events to detect possible exploits and violations, the replay is used to verify if those events are false positives \cite{Shalabi2018RecordReplayAA}. It is also used to analyze time-of-check to time-of-use race conditions \cite{Dunlap2008ExecutionRO} or to determine if systems were previously exploited once zero-day attacks are discovered \cite{Joshi2005DetectingPA}. RnR can be done at different abstraction layers, however, to the best of our knowledge we are the first to record and replay the VMM history in hardware-assisted virtualization solutions.

\section{Discussion and Future Work}
\label{sec:discussion}

Despite positive results obtained by using the proposed record and replay framework, we briefly discuss limitations and avenues of further improvement.

\revision{\textbf{\revision{Memory-related} VM seeds effectiveness}. \toolname{} framework does not replay accurately some \textit{VM behaviors} as discussed in the \subsectionautorefname{}~\ref{subsec:arch_replayig}. Specifically, the recording mechanism deliberately does not store guest VM memory areas touched during VM exit handling. These areas can be related to \revision{Memory-mapped IO (MMIO) and Port-mapped IO (PMIO)} operations between the guest VM and virtualized devices. The behavior of some virtualized devices would be neglected by the current version of IRIS when they use DMA areas, which are beyond the VM exit mechanism.
We plan to explore a way to both record efficiently guest VM memory areas and replay accurate the recorded VM seeds, in terms of code coverage. For example, we can exploit the Extended Page Table (EPT) {\cite{intel2009intel}} approach provided by Intel processors to virtualize the memory resource in hardware-assisted virtualization. We could only record accessed memory areas during the workload execution at the hypervisor level, and link them to the information already gathered by the \textit{IRIS} framework with the VMCS. Moreover, we can exploit dedicated utility functions implemented by a hypervisor, to read and write guest memory areas. For example, the Xen {\lmttfont hvm\_copy\_from\_guest()}/{\lmttfont hvm\_copy\_to\_guest()} routines are used in the {\lmttfont copy\_from\_user\_hvm()}/{\lmttfont copy\_to\_user\_hvm()} core routines as guest memory accessors. Anyway, this software-based mechanism would be less performing compared to the EPT mechanism, which is hardware-based.}


\textbf{Replaying efficiency}. According to the evaluation part, \toolname{} replays \textit{VM seeds} with an efficiency that settles around half of the ideal replaying throughput of $50000$ VM exits/s in our testbed. However, at the current state, \toolname{} does not include any form of optimization for \textit{VM seeds} submission.
\revision{In fact, VM seeds are submitted one by one; the replay mechanism consumes one \textit{VM seed} and waits for the next. Typically, a fuzzer, generates a large set of fuzzing inputs (i.e. \textit{VM seeds} in our case), applying concurrently multiple mutations to the initial seed. Submitting a \textit{VM seeds} in batch, or implementing buffering mechanisms to continuously submit \textit{VM seeds} as they are generated, could increase the overall replay throughput.} In the next releases of \toolname{}, we plan to implement these architectural optimizations.

\textbf{Code coverage}. The code coverage is a paramount metric to guide a fuzzer in interesting points of target source code. The current implementation of \toolname{} leverages a software-based approach like \textit{gcov} \cite{gcov} to retrieve such code coverage data independently of the CPU architecture. Other hardware-based mechanisms, like \textit{Intel Processor Trace} (Intel PT) \cite{intel2009intel}, allows recording complete control flow with low-performance overhead while not modifying the target hypervisor. We plan to experiment \textit{Intel PT} (see Chap. 35 in \cite{intel2009intel}) in \toolname{} to make feasible an efficient coverage-guided fuzzer. However, Intel PT can not be used for a hypervisor that does not target Intel VT-x extensions. \revision{Also, we plan to use \textit{gcov} {\lmttfont flush()} routine to push the coverage data periodically with no need of manual retrieving.} 

{\revision{Besides, code coverage should be accompanied by other metrics to cover critical areas that describe the VM (mis)behavior during execution. To mitigate this point, IRIS allows monitoring of VMCS \{field, value\} pairs, which are peculiar to the hardware-assisted hypervisor solutions. However, it is necessary to deepen the understanding of which other relevant metrics could be used to enable the proper assessment of hardware-assisted virtualization.}

\textbf{Fuzzing}. In this study, we only provided a proof-of-concept fuzzer based upon a record/replay framework, as an example of an assessment solution that addresses hardware-assisted virtualization. However, the simpler mutation rules adopted do not cover the complex fuzzing logic that is adopted by current state-of-the-art fuzzers. We plan to perform a thorough fuzzing experimentation, exploiting the findings provided in this study, to develop a fuzzer aimed at discovering vulnerabilities for hardware-assisted hypervisors. 


\textbf{Portability}. \toolname{} framework is currently implemented for Xen hypervisor and Intel VT-x virtualization extensions. The proposed design can be easily ported to hypervisors that support Intel VT-x (e.g., KVM \cite{kivity07kvm}) since the logic of VM exit handling is well-defined. Regarding the target CPU architecture, different vendors provide their own virtualization extensions. AMD SVM \cite{amd_svm} defines the \textit{Virtual Memory Control Block} (VMCB) data structure, which holds information for the hypervisor and the guest similarly to the VMCS. AMV SVM introduces the \textit{world switch} to indicate the context changes between the hypervisor and guests, and the VMCB for a guest has settings that determine what actions cause the guest to exit to host. ARM virtualization extensions (VHE) \cite{dall2016arm} are centered around privilege levels, also called \textit{exception levels}. ARM includes running user and kernel at EL0 and EL1 levels, respectively, and adds a new CPU privilege level (EL2) to run hypervisor code. A hypervisor running in EL2 can configure the hardware to support VMs, which run in EL0 and EL1 levels. VMs execute normally until some operation (e.g., sensitive instructions) requires the intervention of the hypervisor, leading to traps into EL2 giving control to the hypervisor, and implementing a classic context switch between processes. We plan to address the peculiarities of the most popular CPU virtualization extensions in the next releases of \toolname{}.

{\revision{\textbf{Multiprocessor and multicore support}. Multiprocessor support provided by a hypervisor can be a source of non-determinism and security-related concerns. Accordingly, the \toolname{} framework should differently treat Symmetric Multiprocessing (SMP) and Asymmetric Multiprocessing (AMP) architectures, since the hypervisor code, and consequently, the VM exit handler code could be common to all the physical processors or not. At the time of writing, we performed experiments only by focusing on the Xen hypervisor, which currently supports only SMP architecture for the Intel processors. We plan to explore AMP support of popular hypervisors, like the Xen version for ARM-based platforms. Regarding the issue of having multicore-powered VMs, the current version of IRIS can record and replay VM behaviors according to the VMCS structure provided by Intel VT-x, which is created for each virtual CPU. Thus, the \toolname{} framework can record/replay different flows of vCPU behaviors in the same VM. For example, the Xen hypervisor leverages a structure named {\lmttfont struct vcpu} that keeps track of the current vCPU exited towards the hypervisor.}}

\section{Conclusion}
\label{sec:conclusion}

We presented \toolname{}, a framework to record and replay \textit{VM behaviors} of hardware-assisted virtualization to aid the development of effective fuzzing solutions. The main idea behind \toolname{} framework is to use key abstractions used during switching into the hypervisor code, which is mainly triggered by \textit{VM exiting} mechanism in hardware-assisted solutions. \toolname{} framework allows building \textit{VM seeds} automatically and submitting them in an \revision{accurate} and efficient manner. Indeed, we provided a thorough experimental analysis demonstrating that \toolname{} framework is able \revision{to \textit{record} (learn) sequences of inputs (i.e., VM seeds) from the real guest execution and \textit{replay} them \textit{as-is} efficiently (in the order of $~20K$ \textit{VM seeds} per second) to reach valid and complex VM states. Finally, we use such VM seeds as valid seeds for fuzzing.} We release a proof-of-concept fuzzer to demonstrate the potential of the record and replay approach against the Xen hypervisor, as a first step for assessing hardware-assisted virtualization. 



\IEEEtriggeratref{32}

\bibliographystyle{IEEEtran}
\bibliography{bibliography}

\end{document}